\documentclass[12pt]{iopart}

\usepackage[utf8]{inputenc} 
\usepackage[T1]{fontenc}    
\usepackage{hyperref}       
\usepackage{url}            
\usepackage{booktabs}       
\usepackage{nicefrac}       
\usepackage{microtype}      
\usepackage[ruled,vlined]{algorithm2e}
\usepackage{graphicx}
\usepackage{color}
\usepackage[table]{xcolor}

\usepackage{amsmath,amssymb,amsfonts}

\newcommand{\RNum}[1]{\uppercase\expandafter{\romannumeral #1\relax}}

\newcommand{\x}{\mathbf{x}}

\usepackage{lineno}

\definecolor{cherryblossompink}{rgb}{1.0, 0.72, 0.77}
\definecolor{babyblueeyes}{rgb}{0.63, 0.79, 0.95}
\definecolor{pastelorange}{rgb}{1.0, 0.7, 0.28}
\definecolor{caribbeangreen}{rgb}{0.0, 0.8, 0.6}

\begin{document}

\title{Efficient training of artificial neural network
surrogates for a collisional-radiative model through adaptive parameter space sampling}

\author{Nathan A. Garland }
\ead{n.garland@griffith.edu.au}
\address{Theoretical Division, Los Alamos National Laboratory, Los Alamos, NM 87545, USA}
\address{Faculty of Science and Engineering, Southern Cross University, Lismore, NSW 2480, Australia}
\address{School of Environment and Science, Griffith University, Nathan, QLD 4111, Australia\footnote{Present address}}

\author{Romit Maulik}
\address{Mathematics and Computer Science Division,	Argonne National Laboratory, Lemont, IL 60439}

\author{Qi Tang}
\address{Theoretical Division, Los Alamos National Laboratory, Los Alamos, NM 87545, USA}

\author{Xian-Zhu Tang}
\address{Theoretical Division, Los Alamos National Laboratory, Los Alamos, NM 87545, USA}

\author{Prasanna Balaprakash}
\address{Argonne Leadership Computing Facility,	Argonne National Laboratory, Lemont, IL 60439}
\address{Mathematics and Computer Science Division,	Argonne National Laboratory, Lemont, IL 60439}

\vspace{10pt}

%
%
%
%
%

\begin{abstract}
	Reliable plasma transport modeling for magnetic confinement fusion depends on
	accurately resolving the ion charge state distribution and radiative power
	losses of the plasma. These
	quantities can be obtained from solutions of a collisional-radiative (CR) model at each time step within a plasma transport simulation. However, even compact, approximate CR models can be computationally onerous to
	evaluate, and in-situ evaluations of these models within a coupled plasma transport code can lead to a rigid
	bottleneck. A way to bypass this bottleneck is to deploy artificial neural network surrogates for
	rapid evaluations of the necessary plasma quantities. However, one issue
	with training an accurate artificial neural network surrogate is the reliance
	on a sufficiently large and representative  data set for both training and validation, which can be time-consuming to generate. In this study we further explore a data-driven active learning and training scheme to allow autonomous adaptive sampling
	of the problem parameter space that ensures a sufficiently large and meaningful set of training data assembled for the surrogate training. Our numerical experiments show that in order to produce a comparably accurate CR surrogate, the proposed approach  requires a total number of data samples that is an order-of-magnitude smaller than a conventional approach.\\
	LA-UR-21-31682. Approved for public release; distribution is unlimited.
\end{abstract}


\hspace{.64in}


\newpage
\tableofcontents
\newpage

\section{Introduction}
The construction of ITER, and its components, is now well underway and being performed over the world, with the final assembly and operation to take place in Southern France. ITER will be the world's largest tokamak, which is a device that magnetically confines a hot plasma in the shape of a torus, and it has the goal of demonstrating a path to controlled fusion energy~\cite{editors_chapter_1999,shimada_chapter_2007}. With anticipated experiments to begin in the mid 2020s, there is currently a significant
worldwide research effort to better understand how to efficiently, and safely, operate ITER. Modeling and simulations are a major tool for such studies, with many plasma simulation codes that integrate multiple physics modules to better describe the wide array of physics phenomena taking place within the plasmas~\cite{ferraro_3d_2018,guo_phase-space_2017,mcdevitt_avalanche_2019,hollmann_study_2020,whyte_disruption_2003}. One key component in fusion plasma simulations is to accurately describe the collisional-radiative balance of the plasma, in order to determine impurity ion populations, and the radiation being emitted by the excited ion populations within the plasma at a given time and location~\cite{pigarovphysicslettersa1996,garland_impact_2020,fourniernucl.fusion2000,ralchenko2016modern}.

This information is crucial to accurate modeling of a device like ITER given that impurities, other than the hydrogen-like fuel (H, D, T), will be an inevitable presence in the plasma.
Situations of interest include alpha particles (helium nuclei, that may recombine to He$^+$ or atomic He) produced via the D-T fusion reaction~\cite{editors_chapter_1999}
\begin{eqnarray*}
 ^2_1 \mathrm{D} + ^3_1 \mathrm{T} \rightarrow ^4_2 \mathrm{He} + \mathrm{n}^0,
\end{eqnarray*}
containment vessel impurities, such as beryllium or tungsten sputtering into the plasma~\cite{matthews_melt_2016}, or purposefully injected impurities, such as nitrogen for steady-state plasma power exhaust~\cite{kallenbach_impurity_2013,raj_effects_2020} or neon for tokamak disruption mitigation~\cite{whyte_disruption_2003,strait_progress_2019}, both through enhanced radiative cooling of the plasma.

Presently, a major threat to ITER's safe operation comes from major disruptions~\cite{strait_progress_2019,lehnen_disruptions_2015,lehnen_executive_2018}. A disruption is a sudden termination of the plasma discharge, and can lead to extreme heat load deposition to the reactor walls, excessive electromagnetic forces exerted on the reactor structure, and relativistic electron beam formation that could penetrate through reactor walls~\cite{lehnen_disruptions_2015,lehnen_executive_2018}. In any of these scenarios, the end result could be unacceptable hardware damage that would cost exorbitant time and financial resources to repair. Presently, impurity injection is the reference ITER disruption mitigation strategy, with neon being the top candidate for injection into the hydrogenic plasma~\cite{lehnen_executive_2018}.

Given the need of modeling complex fusion plasma discharges, it is paramount to have accurate ion charge state distribution and radiative power loss rate  when simulating the effects of impurities in the plasma. A general approach to determining these quantities is through collisional-radiative (CR) modeling, which will be further discussed in Section~\ref{sec:CR}. While solution of a CR model will provide accurate plasma properties as input to subsequent coupled stages of a plasma simulation code, the computational time of CR models can be onerous, even when considering relatively simplified models.

Relatively compact approximate CR models, with state vectors $\mathcal{O}(10^1 - 10^2)$ such as that used in this present work~\cite{chunghighenergydensityphysics2005,garland_impact_2020}, can take up to dozens of seconds to solve. The state-of-the-art, fine structure $\mathcal{O}(10^6 - 10^9)$ CR models can take minutes or hours to obtain a solution~\cite{ralchenko2016modern,fontesj.phys.b:at.mol.opt.phys.2015}. Presently, in order to obtain these quantities within plasma transport codes the fusion community has resorted to simplified models, in so-called coronal equilibrium or coronal-like limits, that can limit the fidelity of physics retained in the model~\cite{ralchenko2016modern,bauche2015atomic,summersaipconferenceproceedings2011,whyteproc.24theur.conf.control.fusionplasmaphys.berchtesgad.1997}.

In this study we explore an alternative approach to allow rapid evaluation of necessary CR quantities in the form of an artificial neural network (ANN) surrogate, without sacrificing the fidelity of  the CR model or severely limiting the domain of reasonable input parameters that may be required for model evaluations. Once trained,
owing to fast evaluation time of an ANN-based surrogate, it is expected that coupling the ANN-based surrogate within a plasma simulation framework, in lieu of a CR model solve, will allow accurate atomic and plasma physics information to be passed to the plasma simulation platform, avoiding the computational burden of numerically solving an unnecessarily large nonlinear problem whenever the plasma density is updated, either at each time step in a loose coupling code or at each iteration between multiple models in a tight coupling code.

A standard approach for ANN training in deep learning is to assemble a sufficiently large set of data for both training and validation. This generally enables robust learning of the targeted mappings between input variables and output variables.
One significant challenge when applying such a procedure to the CR surrogate is that the higher fidelity of the atomic energy level description employed in certain CR models, as well as the atomic number of the impurity being considered, may lead to the forward pass evaluation being computationally expensive, as noted previously.
Thus, it is expected that training data generation could become a very time-consuming part when constructing a sufficiently accurate surrogate for a CR model.

To address this potential pitfall, starting from Section~\ref{sec:results1}, we present steps of a procedure including (1) adaptively sample over the parameter space to autonomously determine where additional sample points are required, (2) generate those samples, and (3) retrain the ANN surrogate with possible data re-generation until a pre-determined accuracy criterion is met. The above steps lead to an efficient method that can produce a physically representative set of training data, train the ANN surrogate, and ultimately yield an accurate surrogate, which is demonstrated in Section~\ref{sec:results2} to outperform the one trained on an order of magnitude more data obtained by the standard Latin hypercube sampling.

Finally, it is interesting to note that deep learning applied to plasma physics problems has a rich history. Dating back to the wave of deep learning in the 1990s \cite{van_milligen_neural_1995,allen_design_1992,morgan_feasibility_1991}, through to more recent applications~\cite{kates-harbeck_predicting_2019,rea_progress_2020,maulik2020recurrent,wang_deep_2020,humphreys_advancing_2020}, we believe there is still much to be gained from continued development of plasma physics and deep learning capabilities hand-in-hand. As such, in this study we hope to empower multiphysics fusion plasma modeling of important problems, such as tokamak disruption mitigation, through the application of active deep learning methods to  an ANN-based CR surrogate.

\section{Collisional-radiative modeling and its application }\label{sec:CR}


Accurate modeling of fusion plasmas with added impurities is indispensable to understand the physics for tokamak disruption mitigation~\cite{lehnen_disruptions_2015}. In the simplest scenario, a typical tokamak fusion plasma will be created from hydrogen (deuterium/tritium) gas, which is subject to tremendous amounts of heating to strip electrons from atoms and produce a fully ionized plasma of electrons, positively charged ions with a charge of +1, and possibly a minute fraction of neutral atoms near the boundary.

For the sake of simplicity, in the remainder of this study we will just refer to a deuterium plasma as intended for eventual operation of ITER, where the number density of deuterium species is $n_D$, often on the order of $10^{13}-10^{15}$ $\mathrm{m}^{-3}$ depending on a given mitigation scenario. Densities of the singly charged ion and atom may be referred to as $n^{+1}_D, \ n^0_D$, where their sum is equal to the total density of deuterium species i.e. $n_D = n^{+1}_D+n^0_D$. Additional descriptive variables of a fusion relevant discharge are the density of electrons in the plasma, $n_e$, and the average temperature of the plasma electrons, $T_e$, which characterizes a Maxwell-Boltzmann equilibrium energy distribution,
\begin{eqnarray}\label{f_M}
    f_M(E) = \frac{2}{T_e^{3/2}}\sqrt{\frac{E}{\pi}}\exp \left( \frac{-E}{T_e}\right),
\end{eqnarray}
where $E$ is electron energy in electronvolts (eV) [1 eV $\approx 1.602\times 10^{-19}$ J], and $T_e$ is the electron temperature in eV. Typically fusion plasma temperatures are on the order of $10^3 - 10^4$ eV.

When a discharge is composed of purely hydrogenic atoms or ions~(with atomic number $Z=1$), the atomic physics considerations are generally much simpler, and one may defer to lower-fidelity and cheap models to describe the system with reasonable accuracy~\cite{bauche2015atomic}. However, in recent times the fusion community has been trialing the addition of impurity gases into the plasma to greatly enhance radiative cooling, and thus prevent the deposition of intolerably large levels of heat against the device's first walls that contain the plasma~\cite{lehnen_disruptions_2015,harvey_runaway_2000,hollmann_study_2020}. For the purposes of this study, we can characterize the amount of impurity gas added via a number density, $n_I$. Typical gases in these scenarios include helium~($Z=2$), nitrogen~($Z=7$), neon~($Z=10$), and argon~($Z=18$) and  knowledge of the populations of these atoms' charge states is crucial to accurately modeling a discharge.

We note that a further consideration of adding the aforementioned impurities to a high temperature hydrogenic plasma around $10^3 - 10^4$ eV is that it will result in cooling of the plasma down to $10^0 - 10^1$ eV, and thus a wide range of electron temperatures, $T_e$, must be considered.

Therefore, given some impurity of atomic number $Z$ and density $n_I$, added to a plasma with deuterium density $n_D$ described by an electron temperature $T_e$, the modeling goal is to determine the population densities of the deuterium atom and ion, as well as the impurity atom and ions, in an equilibrium plasma environment (i.e. spatially and temporally homogeneous). Given the wide range of possible ion states that can occur in a plasma, we henceforth will refer to any generic ion level population as $n_j$, and the set of all ion populations, $n_j$, in a given plasma as $\mathbf{n}$. An additional output variable we seek is the total radiative power loss (RPL), $P_{\mathrm{rad}}(\mathbf{n})$, of the plasma given a certain $T_e$ and $(n_D, n_I)$. This scalar quantity is a function of $\mathbf{n}$ and describes contributions to radiation being emitted from radiatively decaying excited ion states, electrons radiatively recombining with ions, and electron Bremsstrahlung radiation~\cite{ralchenko2016modern,tallents_2018}.

To access the desired populations of all ions in the plasma, $\mathbf{n}$, and the radiation power loss, $P_{\mathrm{rad}}(\mathbf{n})$, a CR model is used to solve for the populations of ions in various excited energy states~\cite{ralchenko2016modern,bates_recombination_1962,mcwhirter_calculation_1963,chunghighenergydensityphysics2005}. A CR model is formulated as a set of
ordinary differential equations (ODEs) for each ion population, $n_j$, that describe transitions into ($i\rightarrow j$) or out of ($j\rightarrow i$), in relation to all other possible ion populations, $n_i$, where $1\leq i,j \leq N$ and $N$ is the total number of ion populations in the state vector, $\mathbf{n}$. This rate problem can be expressed as
\begin{eqnarray}\label{cr-eq1}
    \frac{dn_{j}}{dt} = \sum_{i=0,i\neq j}^N R_{i\rightarrow j}n_i - \left(\sum_{i=0,i\neq j}^N  R_{j\rightarrow i} \right)n_j,
\end{eqnarray}
where each $R_{i\rightarrow j}$ is a sum over the possible collisional and radiative transitions that can occur in a plasma, which depends on $\mathbf{n}$ in general. These include excitation and ionization of atoms/ions via electron impact collisions,  radiative decay of excited states, and radiative recombination of ions~\cite{ralchenko2016modern,tallents_2018}. The rates of these atomic physics events are dependent on the electron energy distribution in the plasma, which can add another degree of difficulty to the problem. As alluded to previously, the Maxwell-Boltzmann distribution (\ref{f_M}) for electron energies is assumed for the purposes of this study, meaning most rates must
be computed for a given plasma temperature, $T_e$.

Given the large number of possible species populations, it is natural to form a nonlinear rate matrix problem
\begin{eqnarray}\label{cr-eq}
    \frac{d\mathbf{n}}{dt} = \mathsf{R}(\mathbf{n})\mathbf{n},
\end{eqnarray}
where $\mathsf{R(\mathbf{n})}$ is the rate matrix composed of elements $R_{i\rightarrow j}$ and $\textbf{n}$ is the state vector previously discussed.

When the CR relaxation timescale to an equilibrium is faster than
plasma transport effects being simulated in a coupled code,
it is common to assume a steady-state scenario to determine plasma properties within such a simulation, i.e., one seeks to solve
\begin{eqnarray}\label{cr-eq-ss}
     \mathsf{R}(\mathbf{n})\mathbf{n} = 0.
\end{eqnarray}

In the implementation of most CR models, a range of excited quantum states for each ion stage are specified, $n_{Z_j}^{\alpha_j},$ where the index $Z_j$ denotes an ion charge state, and $\alpha_j$ denotes the excited state level of the $j$th species in $\mathbf{n}$. Generally, plasma transport models will only require the population of each ion stage, irrelevant of what excited quantum states are occupied, and so during the post-processing of a CR model we will sum over all excited states, $\alpha_j$, to find the population of a given ion stage $Z_j,$
\begin{eqnarray}\label{n_ion_eq}
    n_{Z_j} = \sum_{\alpha_j} n_{Z_j}^{\alpha_j}.
\end{eqnarray}

In the application of a CR model, there are many quantities determined by the solution of the model that have use for immediate interpretation of a plasma and its properties, and also use for coupling within a multiphysics plasma simulation framework, such as a plasma particle transport simulation code. The fundamental output quantities of a CR model are the ion population densities, $n_j$, that compose the state vector of the CR system, $\textbf{n}$. When these ion populations are known, the CR model can then compute the radiation loss, $P_{\mathrm{rad}}(\mathbf{n})$.

An important quantity is the free electron density of the plasma, $n_e$, formed as a result of ionization of atoms
and ions in the plasma. In the later sections we will use this quantity to compare the performance of proposed surrogates. In this work, we are considering a deuterium plasma with some higher $Z$ impurities, so the
electron density is given by
\begin{eqnarray}\label{n_e_eq}
    n_e = n^{+1}_D + \sum^Z_{Z_j=0} Z_j n_{Z_j},
\end{eqnarray}
where $n^{+1}_D$ is the number density of deuterium ions, $n_{Z_j}$ is
the number density of the impurity ion with charge $Z_j$, and $Z$ is the atomic number of the impurity species.

An additional output quantity of a CR model used to characterize a plasma is the average ion charge state, $\bar Z$. A weighted average of
ion charge with ion population density, we can write $\bar Z$ for an impurity species as
\begin{eqnarray}\label{zbar}
    \bar Z = \frac{\sum^Z_{Z_j=0}  Z_j n_{Z_j}}{\sum_{Z_j=0}^{Z} n_{Z_j} }.
\end{eqnarray}

Numerical solutions to \eqref{cr-eq-ss} can be computationally challenging when $\textbf{n}$ is high-dimensional, or $\mathsf{R}$ is nearly singular. Solutions of the nonlinear rate-matrix problem \eqref{cr-eq-ss} are usually performed with a standard linear algebra package. Once $\textbf{n}$ is known, the sum of radiation power loss due to each excited ion state is then computed by the CR model, and $P_{\mathrm{rad}}$ is then passed back to the main execution loop of a plasma model~\cite{hollmann_study_2020,hesslow_generalized_2018,mcdevitt_avalanche_2019}.

Ideally, one would simply solve a CR model at each time step or iteration of a plasma code, however as outlined in the previous section the computational bottleneck this would create is unsatisfactory. As such, in the following sections we present an efficient method to assemble physically meaningful training data to train an effective ANN surrogate. Essentially, such a surrogate will provide the required mapping of input plasma conditions to output quantities normally computed from a CR model in 10s of microseconds, rather than seconds or even longer.



\section{Surrogate modeling strategy and demonstration}\label{sec:results1}

With the forward pass CR model detailed, and the requirements of a potential surrogate outlined, we now detail the formulation
of the surrogate modeling strategy. We begin with consideration of the input and output quantities required for the surrogate
mapping, before detailing the two-tiered surrogate forms and the adaptive sampling strategy.

When considering the input quantity of an impurity species' atomic number, $Z$, we must note the fundamentally different nature of plasma particle interactions with targets of different atomic numbers, $Z$. For example, electron scattering from small targets like hydrogen or helium ($Z=1,\,2$) is significantly different to scattering from larger targets such as neon or argon ($Z=10,\,18$). The subsequent impacts this can have on plasma properties are not trivial due to each integer value of $Z$ essentially posing a discrete quantum mechanical scattering problem, which could be very different to the problem for adjacent elements on the periodic table, i.e. $Z-1$ or $Z+1$. As such, we believe it is preferable to choose not to treat $Z$ as a continuous variable to be fed as an input, and rather defer to generation of \emph{fit-for-purpose} surrogates for each required value of $Z$. Given the application of these surrogates is limited to a small number of tokamak fusion relevant elements in the periodic table, the option of fit-for-purpose surrogates specific to each atomic number is a sensible treatment.

Therefore, in this work, for a given impurity under consideration, with atomic number $Z$, any surrogate we seek will provide a map for an input field of: electron temperature, impurity density, deuterium density,  $\{T_e,n_I,n_D\}$, to an output field of the total radiative power loss and ion stage populations for deuterium and the added impurity,
$\{P_{\mathrm{rad}},n_D^0,n_D^{+1},n_I^0,n_I^{+1},...,n_I^{+(Z-1)},n_I^{+Z}\}$.

 To generate sample data for training, we must specify bounds on the input fields. Within typical parameters expected in the application of a CR surrogate to tokamak disruption mitigation, we limit the input field quantities to the domains of $1 \ \mathrm{eV} \leq T_e  \leq 1000 \ \mathrm{eV}$,
$10^{13} \ \mathrm{cm}^{-3} \leq n_I \leq 10^{15} \ \mathrm{cm}^{-3}$, and $10^{13} \ \mathrm{cm}^{-3} \leq n_D \leq 10^{15} \ \mathrm{cm}^{-3}$. In order to regularize the input field
being mapped to output, each variable is normalized to the domain $[0,1]$ via
\begin{align}\label{eq:scale_to_1}
    \hat x = \frac{ x - \min(x)}{ \max(x) - \min(x)},
\end{align}
where $x$ and $\hat x$ are raw and normalized quantities.
We choose this input scaling to allow uniform sampling over $[0,1]$ as we have no \textit{a priori} expectation of possible input quantity values. For example, essentially any one value of deuterium and/or added impurity atom densities composing the plasma is just as likely as another.

The ion populations in the output field are normalized against the input impurity number density, $n_I$, and deuterium density, $n_D$, such that we
assume to be dealing with fractional populations satisfying
\begin{align}
    n_D^0 + n_D^{+1} = 1,\\
    n_I^0 + n_I^{+1}+ \ ... \ + n_I^{+(Z-1)} +n_I^{+Z} = 1.
\end{align}

Given that $P_{\mathrm{rad}}$ may vary by up to ten orders of magnitude on a logarithmic scale, depending on multiple input factors, we choose
to operate with $\log_{10}(P_{\mathrm{rad}})$ in order to bring this element of the output field to be on a similar order
of magnitude as the fractional ion populations.

In the following section, we evaluate the construction of surrogate models for our CR dataset so that crucial plasma quantities may be obtained using inexpensive ANN evaluations. Notably, our goal is to not just obtain a cheap surrogate but to also balance the offline cost of representative data set generation for training this surrogate. This is obtained via an adaptive sampling explained below. To arrive at a final, trained ANN surrogate we use two distinct types of surrogates in this research: a low fidelity (LF) and a high fidelity (HF) surrogate. The two models differ in the ability to accurately represent the CR input-output relationship and are used for two distinct purposes in our surrogate development campaign.

\subsection{Low fidelity surrogate: Random forest regressor}

The first of the two types of surrogate models employed in our work is a low fidelity random forest regressor (RFR) \cite{breiman2001random}. RFR is an ensemble machine learning technique popularly used for regression and classification tasks \cite{hastie2005elements}. It is built around the utilization of many decision trees (thus leading to the `forest' terminology). A decision tree is an efficient algorithm for splitting a data set according to its various independent variables that leads to a branch-like structure with each `node' corresponding to a splitting based on one variable. The decision to split according to one independent variable as opposed to the others is taken by measuring the effectiveness of the split through metrics such as impurity or variance reduction in the split.  For a regression task, impurity is defined as
\begin{align}
\label{Eq1}
{H\left(Q_m\right)=\frac{1}{N_{m}} \sum_{i=0}^{N_m}\left(y_{i}-\overline{y}_{m}\right)^{2}},
\end{align}
where $\overline{y}_m$ is the mean of the dependent variable $y_i,$
\begin{align}
    \overline{y}_{m}=\frac{1}{N_{m}} \sum_{i=0}^{N_m} y_{i},
\end{align}
    and  $N_m$ is the number of data points at a particular node $m$. The set $Q_m$ represents the data that resides at a node. The total impurity at this node may then be expressed as
\begin{align}
\label{Eq4}
    G(Q_m, \theta)=\frac{n_{\rm left}}{N_{m}} H\left(Q_{\rm left}(\theta)\right)+\frac{n_{\text {right}}}{N_{m}} H\left(Q_{\text {right}}(\theta)\right),
\end{align}
where $n_{\rm left}$ and $n_{\rm right}$ correspond to the number of children data points in the left and right branches arising from the node $m$ and
\begin{align}
    \begin{gathered}
    Q_{\rm left}(\theta) = (x, y) | x_{j} <= t_{m}, \\
    Q_{\rm right}(\theta) = (x, y) | x_{j} > t_{m}
    \end{gathered}
\end{align}
are the left and right split data sets respectively. Note how the splitting of $Q_m$ depends on an independent variable $x_j$ and a threshold $t_m$ at each node. The choice of $t_m$ is generally given by the median value of the attribute $x_j,$ i.e., the left data set is comprised of all samples with independent variable $x_j$ less than or equal to $t_m$. The decision tree splits the data $Q_m$ in a manner that minimizes $G(Q_m, \theta)$ by choosing an optimal dimension $j$ for the independent variable. This branching is performed recursively for every node until certain user-defined criteria are met such as $N_{m} \le N_{\text{tol}}$ at which point the node is denoted a leaf. A branching may also be terminated if a certain maximum depth is reached. A new data point (i.e., an unseen sample) can then follow the branching trajectory (by tracking the order of splits by $j$ and Equation \ref{Eq4} and placement into left or right branches by $t_m$) and reach a leaf. In case a $N_{\text{tol}}$ criterion is set, the prediction of the decision tree for a sample is the average prediction value of the dependent variables at this leaf.


RFRs utilize ensembles of decision trees by selecting random subsets of the total data for each tree to obtain a consensus on regression predictions. Each tree utilizes a random subset of the data through sampling with replacement and may therefore obtain a completely different branching structure depending on the distribution of that particular data set. The generation of multiple decision trees as estimators also encourages generalization. The underlying mechanism of our adaptive sampling strategy uses the ensemble property of the RFR wherein each tree in the forest may be queried for a prediction of a scalar metric, to estimate prediction variances in the multidimensional parameter space. If the different trees in an RFR show a wide range of predictions (i.e. there is widespread disagreement in the predictions of trees in a forest) in some region of the finely sampled space, we run full-order CR evaluations in that region, append this to our master training data set and retrain the RFR as well as the HF surrogate. This allows us to adaptively sample higher dimensional spaces which may be challenging to cover due to the curse of dimensionality.

\subsection{High fidelity surrogate: Artificial neural network}

Given a LF surrogate used to rapidly inform re-sampling, the second type of surrogate employed in our work is a high fidelity ANN that may act as a universal approximator \cite{hornik1989multilayer} trained by back-propagation \cite{hecht1992theory}. One technique to obtain a high-fidelity function approximation is through the use of a multilayered perceptron (MLP) architecture, which is a subclass of feedforward artificial neural network. A general MLP consists of several neurons arranged in multiple layers. These layers consist of one input and one output layer along with several hidden layers. Each layer (with the exception of an input layer) represents a linear operation followed by a nonlinear activation that allows for great flexibility in representing complicated nonlinear mappings. This may be expressed as
\begin{align}
    \mathcal{L}^{l}\left(\boldsymbol{t}^{l-1}\right):=\boldsymbol{w}^{l} \boldsymbol{t}^{l-1}+\boldsymbol{b}^{l},
\end{align}
where $\boldsymbol{t}^{l-1}$ is the output of the previous layer, and $\boldsymbol{w}^l,\boldsymbol{b}^l$ are the weights and biases associated with that layer. The output, $\boldsymbol{t}^l$, for each layer may then be transformed by a nonlinear activation, $\eta$, such as rectified linear activation:
\begin{align}
    \eta(a) = \text{ReLU} (a) = \max(a,0).
\end{align}
For our experiments, the inputs $\boldsymbol{t}^0$ are in $\mathbb{R}^d$ (i.e., $d$ is the number of inputs) and the outputs $\boldsymbol{t}^K$ are in $\mathbb{R}^k$ (i.e., $k$ is the number of outputs). The final map is given by
\begin{align}
    F: \mathbb{R}^{d} \mapsto \mathbb{R}^{k}, \quad \boldsymbol{t}^0 \mapsto \boldsymbol{t}^K = F(\boldsymbol{t}^0 ;(\boldsymbol{w}, \boldsymbol{b})),
\end{align}
where
\begin{align}
    F(\boldsymbol{t}^0 ; \boldsymbol{w},\boldsymbol{b})=\eta^{K}\left(\mathcal{L}^{K} \circ \eta^{K-1} \circ \mathcal{L}^{K-1} \circ \ldots \circ \eta^{1} \circ \mathcal{L}^{1}\right)(\boldsymbol{t}^0)
\end{align}
is a complete representation of the neural network, and $\mathbf{w}$ and $\mathbf{b}$ are a collection of all the weights and the biases of the neural network. 
These weights and biases, lumped together as $\phi = \{\mathbf{w}, \mathbf{b} \}$, are trainable parameters of our map, which can be optimized by examples obtained from a training set. The supervised learning framework requires for this set to have examples of inputs in $\mathbb{R}^{N_{ip}}$ and their corresponding outputs $\mathbb{R}^{N_{op}}$.
The corresponding cost function is given by
\begin{align}
    \mathcal{C} = \frac{1}{|T|} \sum_{(\boldsymbol{\theta}, \tilde{\x}) \in T}\left\|\tilde{\x} - F(\boldsymbol{\theta};\phi)\right\|^{2}
\end{align}
where $|T|$ indicates the cardinality of the training data set, with input, $\theta$, and output fields, $\tilde{\x}$, given by
\begin{align}
    T=\left\{\left(\boldsymbol{\theta}_i, \tilde{\x}_i\right): \tilde{\x}_i=f\left(\boldsymbol{\theta}_i\right)\right\}.
\end{align}
Here $f\left(\boldsymbol{\theta}_i\right)$ are examples of the true targets obtained from the adaptively generated training data set. The trained MLP may be used for uniformly approximating any continuous function on compact domains \cite{cybenko1989approximation,barron1993universal}, provided $\eta(x)$ is not polynomial in nature.

\subsection{Training with adaptive sampling}\label{sec:hyperp}

In this section, we introduce a methodology to iteratively use two surrogate models - the RFR and ANN for reducing overall computational costs and balancing accuracy with data generation costs. The high-level idea is to use the RFR to identify regions where predictions of an ANN regressor could be improved by expanding the training data set. By coupling the two surrogates for this approach - accurate surrogate models may be obtained that leverage the expressiveness of the ANN but rely on the ease of training of the RFR for identifying a superior dataset. Note that one may also use ANNs that provide uncertainty estimates for identifying regions of high variance for adaptive sampling. However, training such networks is generally more expensive than regular networks and imprecise in the limit of small data. We note, however, that by choosing surrogates from two different classes, the RFR and ANN, we utilize an assumption (i.e., an inductive bias) that the predictive variance from the predictions of different decision trees imply an inability to learn a consistent function approximation in an ensemble sense. It follows then, that the predictive variance is an appropriate region to target for sampling to improve any surrogate model. Such inductive biases have proven effective for various function approximation tasks \cite{balaprakash2018deephyper}.

To initialize this adaptive algorithm, a first (relatively coarse) sampling of the input space is constructed to train our RFR following which the adaptive sampling is performed iteratively. The end of an adaptive sampling iteration is accompanied by a retraining of the HF surrogate on the expanded training data set. The validation performance of the HF surrogate is recorded for the purpose of \textit{a posteriori} model selection. We note that the sampling at each iteration of the LF surrogate (and at the very start of the adaptive training data augmentation) utilizes the Latin Hypercube sampling (LHS) method for experimental design~\cite{LHScode}. We note that a choice of scalar metric for the LF surrogate must be made in order to guide resampling. In this study we trial physical quantities in this role, one being the average charge state of the impurity ions $\bar Z $ and the other the total radiative power loss, $P_{\mathrm{rad}}$. Both scalar quantities are physically meaningful and important quantities, but represent different physical aspects of a plasma. The total radiative loss, $P_{\mathrm{rad}}$, is the first element of the output layer, while $\bar Z $ is computed as a weighted mean via \eqref{zbar}. Given \textit{a priori} neither of them trumps the other in terms of which should be the more useful guide, we trial both.

The hyperparameters that need to be selected \textit{a priori} for this framework include the number of samples required in the first training iteration, the number of new samples for each data set augmentation, the total computational budget for training the surrogate, and the traditional hyperparameters of the MLP (e.g., the learning rate, early stopping criterion, architecture) and the RFR (the number of trees, maximum depth of the forest). Our sampling strategy is outlined in Algorithm \ref{ASS} where the hyperparameters of the associated components are as follows: $N_{\rm init}=50$ initial samples, $N_{\rm resample}=0.1N_{\rm samples}$ re-samples per data set augmentation iteration given a current number of acquired samples $N_{\rm samples}$, and a budget of $N_{\rm budget}=1950$ samples to make 2000 samples the maximum possible. The MLP was chosen to have two hidden layers, 50 neurons wide each, between the input and output layers of dimension three and 14 respectively. The ReLU activation function was employed in this network. The learning rate was fixed as $10^{-3}$ with the Adam algorithm \cite{kingma2017adam} used for optimization. An early stopping criterion was set to halt learning after 100 epoch of failing to reduce validation loss. Hyperparameters of the RFR were chosen as 50 trees in the forest, with a maximum depth of eight. The stopping criterion for the overall adaptive sampling framework was set to be when a validation $R^2$ result of 0.95 or greater is obtained.

\begin{algorithm}[]
	\label{ASS}
	\SetAlgoLined
	Generate $N_{\rm init}$ initial samples of forward CR model evaluation\;
	Train HF surrogate\;
	\While{$N_{\rm samples}<N_{\rm budget}$}{
		Train LF surrogate to map input field to scalar metric\;
		Evaluate fine sample space grid using LF surrogate. Sort list by variance\;
		\For{$i$ = $1:(N_{\rm samples}/10)$}{
			Take input field of highest variance LF prediction. Generate new sample\;
		}
		Train HF surrogate on expanded sample set. 	Evaluate surrogate metrics.\;
		\If{$R^2 > R^2_{\rm goal}$}{
			STOP}
	}
	\caption{Adaptive CR surrogate training sampling}
\end{algorithm}



\subsection{Experiments and discussion}
Given the proposed adaptive sampling method, we now examine the performance of the adaptive sampling routine and compare the performance of its resulting surrogate versus a traditional approach for
data generation and training of an ANN. All assessments in this study were conducted with Tensorflow 2.0.0 under Python 3.7.9 on a 2.9GHz Intel Core i9 CPU with 32GB of RAM. In this Section we choose to specify $Z=10$, which corresponds to a deuterium and neon plasma
mixture, a crucial component for planned ITER disruption mitigation strategy~\cite{lehnen_executive_2018,strait_progress_2019}.

\begin{figure}[]
	\centering
	\includegraphics[width=0.9\linewidth]{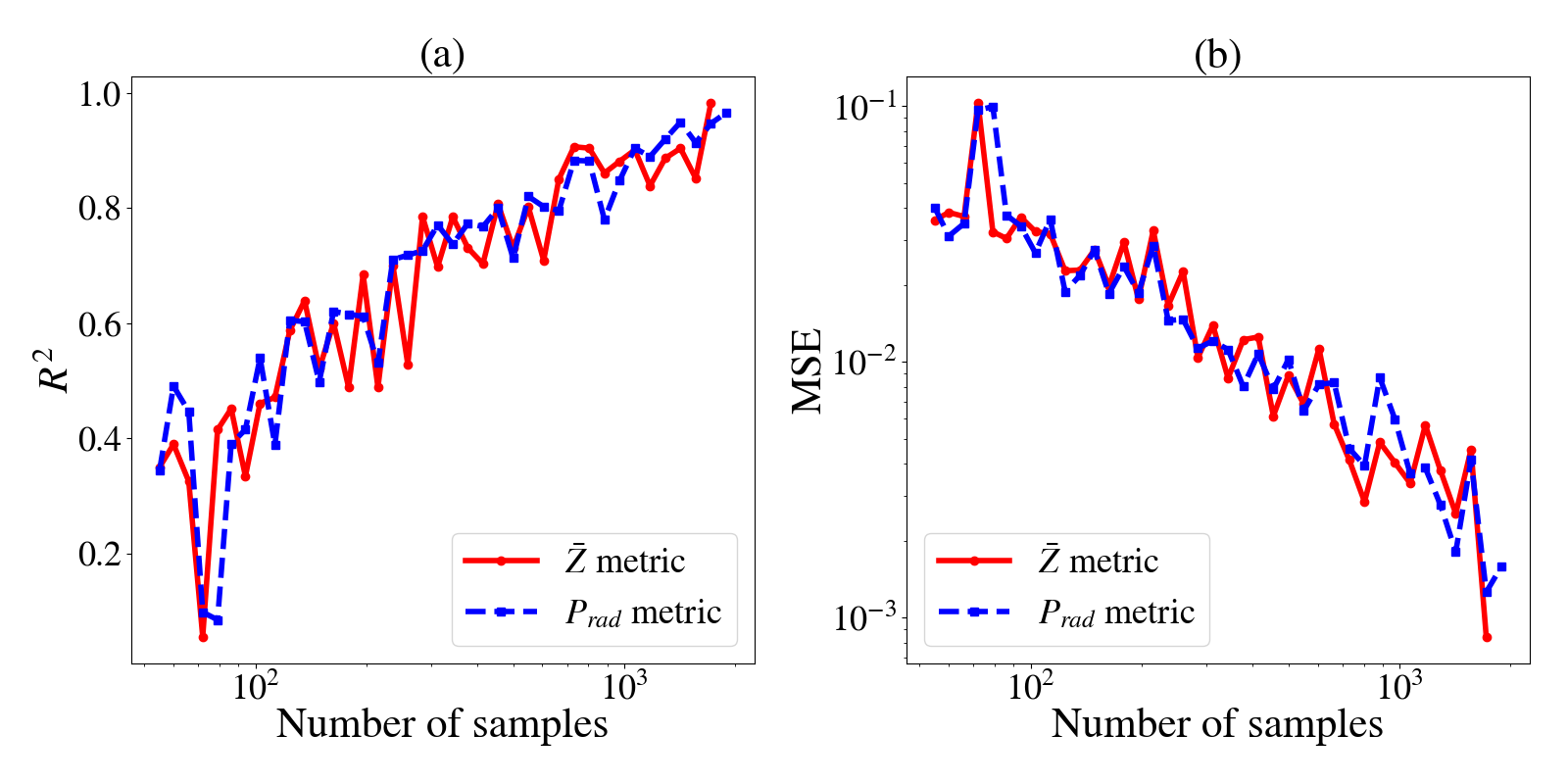}
	\caption{Training validation metrics for (a) $R^2$ and (b) mean-squared error (MSE) of adaptively trained MLP surrogate as a function
	of acquired training samples. Clear, but relatively minor, differences are seen between using either $\bar Z$ or $P_{\mathrm{rad}}$ as the decision metric for the RFR low fidelity surrogate. }
	\label{fig:train_metric}
\end{figure}

Employing the network architectures and hyperparameter choices outlined in the previous section, we now present some representative results from data generation and training exercises to
examine the ability of the adaptive sampling framework when producing an ANN-based surrogate trained on autonomously gathered training data. During the training, we employ two scalar metrics of $\bar Z$ and $P_{\mathrm{rad}}$ in RFR decision making of the adaptive framework. It is found that both cases produced a steady
convergence in validation $R^2$ performance and drop in validation MSE as more samples were acquired, as shown in Figure~\ref{fig:train_metric}. While non-monotonic behavior is indeed seen in both quantities with increasing samples added, we can observe
that these fluctuations follow an alternating prediction-correction behavior, which is an expected result of the designed
sampling routine. Ultimately the adaptively trained MLP surrogates passed the $R^2$ threshold of 0.95 and final surrogates were obtained.
We found
that the performance
difference between $\bar Z$ or $P_{\mathrm{rad}}$ scalar metrics is usually small, however it was observed that the $\bar Z$ option
was generally more reliable in reaching the training goal earlier with fewer forward pass model evaluations required. As an example, for the case
shown in Figure~\ref{fig:train_metric} the $\bar Z$ metric case yielded a validation $R^2$ of 0.983 and MSE of $8.422\times 10^{-4}$, after 1719 samples were acquired in training. The case of the $P_{\mathrm{rad}}$ metric yielded a validation $R^2$ of 0.965 and MSE of $1.587\times 10^{-3}$ using 1890 samples.

\begin{figure}[]
	\centering
	\includegraphics[width=0.3\linewidth]{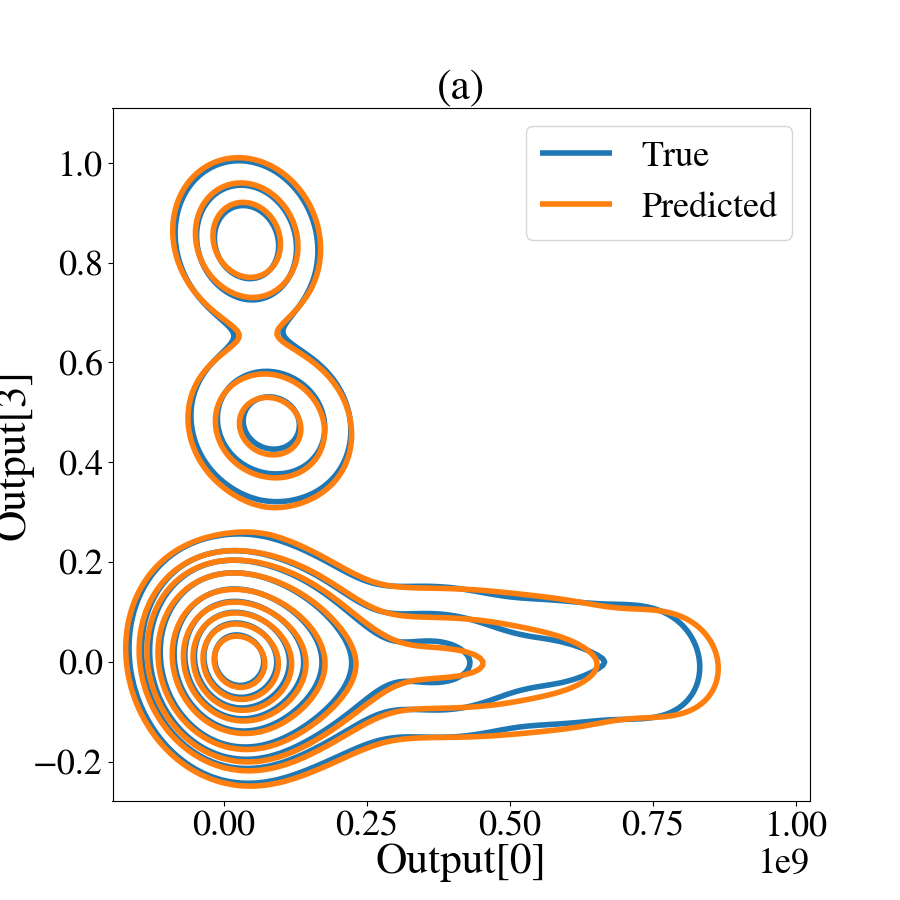}
	\includegraphics[width=0.3\linewidth]{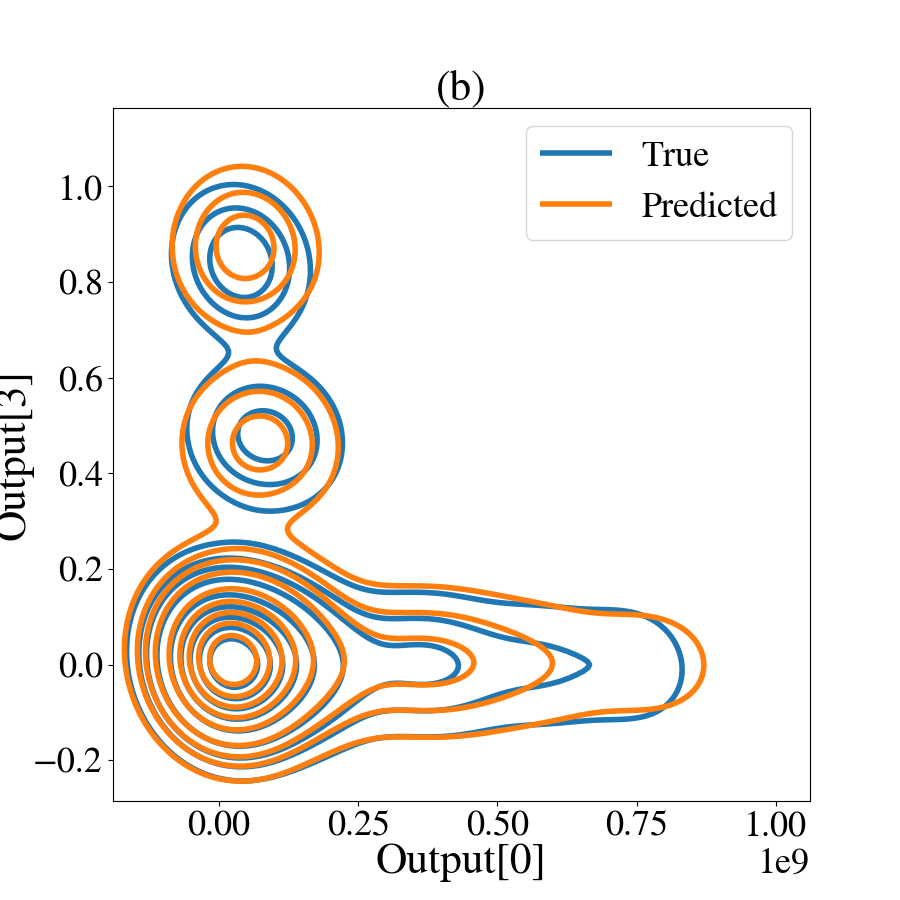}
	\includegraphics[width=0.3\linewidth]{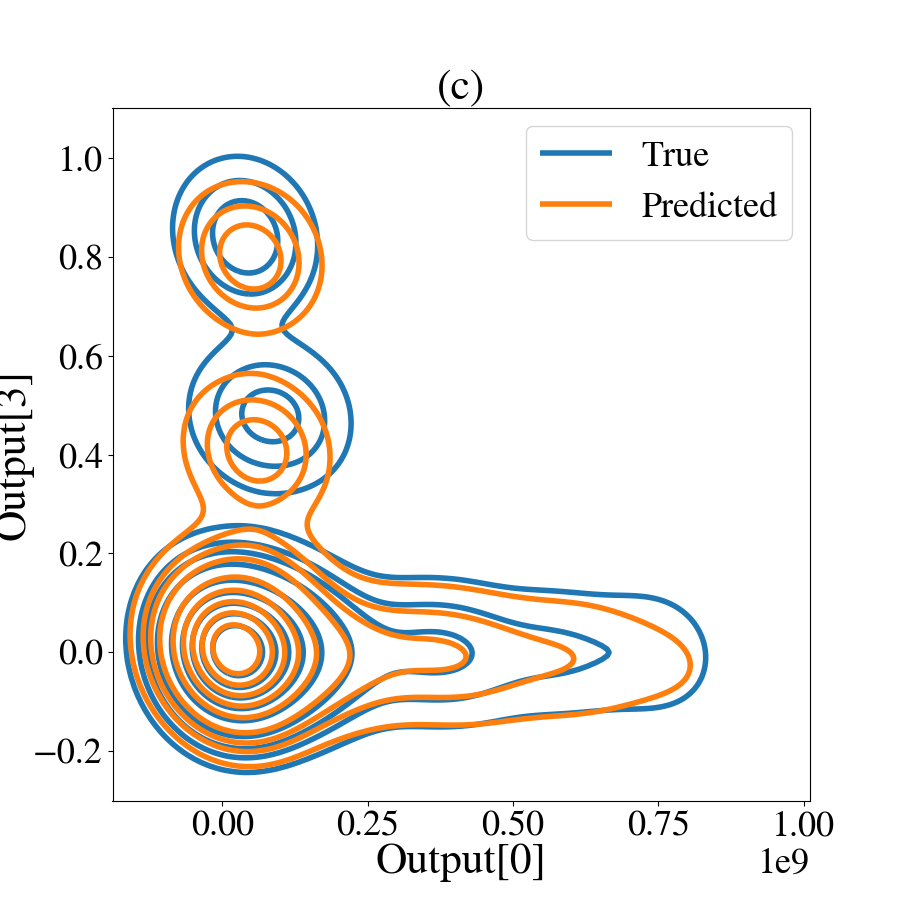}
	\caption{Kernel density estimate plots demonstrating the true and predicted relationship between output field quantities of total radiation $P_{\mathrm{rad}}$ and the population density of the Ne$^{+10}$ ion for (a) MLP conventionally trained on fixed set of 3375 samples, (b) MLP adaptively trained on 1719 samples by using $\bar Z$ as a guiding metric, and (c) MLP adaptively trained on 1890 samples by using $P_{\mathrm{rad}}$ as a guiding metric. }
	\label{fig:PDF}
\end{figure}

As a point of reference for the performance of the adaptive sampling framework, an ANN surrogate was simply trained on a fixed data set of size 3375, which is the result of choosing a space of $15\times15\times15$ samples determined via LHS. The hyperparameters outlined in the previous section were employed in training this surrogate, and a test $R^2$ of 0.999 was obtained against an unseen test set. To demonstrate the ability of this ANN, from this unseen test set, a kernel density estimate (KDE) plot showing the predicted relationship between two output field quantities is shown in Figure~\ref{fig:PDF}.

Testing the adaptively trained MLP surrogates on an unseen data set, held back from training or validation data, we present KDE plots in Figure~\ref{fig:PDF} showing the relationship
between the output field quantities of total radiation $P_{\mathrm{rad}}$ and the population density of the Ne$^{+10}$ ion as a way to demonstrate the performance of predicted quantities versus the true values. Here we see, compared to the ANN trained on a larger training set, the adaptively trained, $\bar Z$-guided surrogate performs reasonably well, with the exception of some translation of the prediction over the two upper islands in the plot. Similarly, the surrogate guided by the $P_{rad}$ metric does not accurately reproduce the upper two islands of the KDE. We note that testing $R^2$ values of 0.982 and 0.967 were obtained for the adaptively trained networks informed by $\bar Z$ and $P_{\mathrm{rad}}$ respectively.

Given the primary motivator of constructing a framework for surrogate training was for rapid in situ evaluations, a timing comparison for 1000 forward pass evaluations of the ANN surrogate and original CR model executable was performed to highlight the computing time savings. The mean surrogate execution time was approximately 10 $\mu$s, while the mean CR model forward pass execution was 2.7 s. While the direct comparison between adaptively trained surrogates discussed in this section and a traditional ANN surrogate trained on a large pre-assembled data set is at best inconclusive, we have demonstrated the success of the sampling and training method as a proof of concept. To extend upon this basis, in the next section, we further refine some of the strategies used in adaptive sampling and demonstrate that a comparable, or often better, ANN surrogate may be produced through automated sampling acquisition.

\section{Improving adaptive sampling in training and data acquisition  }\label{sec:results2}


In the prior section, we examined a brief demonstration of the feasibility in producing a sufficiently accurate surrogate, within the error tolerances forgivable for in-situ coupling in plasma transport codes. Production of this surrogate via an adaptive training-sampling method was shown to be quicker than assembling an arbitrarily large training data set, while still ensuring enough representative sample points were taken over the parameter space. The proof-of-concept example presented in the previous section was achieved while employing a very simple sample acquisition method that gathered from high variance regions for a scalar output of the HF surrogate.

In this section, we present a more effective sample acquisition method that produces an accurate ANN surrogate with very few samples required to match or exceed an ANN surrogate trained on a very large LHS sampled training set. In lieu of the prior choices for the LF surrogate scalar metric, being the average ion charge state or the total radiation power loss, we now use the LF RFR surrogate to predict the validation MSE of the HF surrogate
\begin{eqnarray}
    \mathrm{MSE} = \frac{1}{N} \sum_j^N ( y_j - \hat{y}_j)^2,
\end{eqnarray}
where $N$ is the number of data points in the training set, $y_j$ is the true value of the $j^{th}$ output, and $\hat{y}_j$ is the corresponding output predicted by the ANN output node. 
With a focus to rapidly evaluate an estimation of the HF surrogate's error over a finely grained input parameter space, we now propose an improved training sample acquisition function.

\subsection{Balancing exploitation vs.~exploration}
With the use of a LF surrogate that predicts the sample error of validation data for the HF surrogate, we may utilize acquisition function ideas from Bayesian optimization literature~\cite{agnihotri2020exploring,JMLR:v22:18-220} to adaptively select new points. 
First, we sample a large number of unevaluated input parameters. At each of these locations, the LF surrogate is utilized to compute a mean ($\mu$) and standard-deviation ($\sigma$) estimate for the error. Evaluating points with large values of $\mu$ indicates that this point can potentially result in reduction of the validation error of the HF surrogate, which we denote \emph{exploitation}. Large values of $\sigma$ indicate that the LF surrogate has not adequately learned to predict the error at these locations and therefore these regions may add to the diversity of the overall training data set through a process of \emph{exploration}. Several acquisition
functions have been developed for Bayesian optimization that attempt to balance exploration and exploitation for various settings. In this study we focus on the lower confidence bound, a simple and robust acquisition function defined by
\begin{align}\label{eq:MSE_COST}
    \mathcal{A}_{LCB} (x) = \mu(x) + \lambda \sigma(x),
\end{align}
where the parameter $\lambda \geq 0$ controls the balance of exploration and exploitation. When set to zero, this parameter performs pure exploitation.


In our work we have examined the effects of varying this $\lambda$ parameter to control the balance between acquiring new training data samples from the parameter space region of high mean error (exploitation) and the region of high variance (exploration). As a result, we will present a brief comparison in the subsequent sections outlining the benefits of a balanced approach between exploitation and exploration in assembling a representative and meaningful training data set.

\subsection{Training with new adaptive sampling}
As a point of reference for the performance of the adaptive sampling framework, an ANN surrogate was simply trained on a fixed data set of size 8000, spanning a space of $20\times20\times20$ samples determined via LHS. Here the fixed data set size is increased to 8000 compared to 3375 in the previous section so to raise the bar on the ANN benchmark  that we will compare against. The hyperparameters selected for training the final version of CR surrogate were largely kept the same as the prior training procedure, outlined in the previous section~\ref{sec:hyperp}. Notable changes to these hyperparameters were that we employed
$N_{\rm init}=100$ initial samples, $N_{\rm resample}=0.1N_{\rm samples}$ re-samples per data set augmentation iteration, and a budget of $N_{\rm budget}=7900$ samples to make 8000 samples the maximum possible to match the total number of data points used in generation of a high-resolution data set for training a benchmark ANN surrogate. Additionally, we increased the threshold for stopping criteria of the overall adaptive sampling framework to be a validation $R^2$ result of 0.99 or greater. Further, given the oscillatory
nature of the validation criteria observed in Figure~\ref{fig:train_metric} as a function of increasing samples, we add a requirement that to successfully complete training the goal validation $R^2$ must be passed in sequential iterations of the outer loop. This ensures an improved surrogate is produced, by having to demonstrate robust convergence to the desired criteria and not simply passing once, before potentially failing on the next data set augmentation. Modifications to the sampling-training algorithm are shown in Algorithm~\ref{ASS2}.

\begin{algorithm}[]
	\label{ASS2}
	\SetAlgoLined
	Generate $N_{\rm init}$ initial samples of forward CR model evaluation\;
	Train HF surrogate\;
	\While{$N_{\rm samples}<N_{\rm budget}$}{
		Train LF surrogate to map input field to HF surrogate MSE\;
		Evaluate fine sample space grid using LF surrogate\;
		Evaluate lower confidence bound in (\ref{eq:MSE_COST}) and sort list\;
		\For{$i$ = $1:(N_{\rm samples}/10)$}{
			Take input field of highest variance LF prediction; generate new sample\;
		}
		Train HF surrogate on expanded sample set; 	evaluate surrogate metrics\;
		\eIf{$R^2 > R^2_{\rm goal}$}{
		    Increment pass count $N_{\rm pass}$\;
			\If{$N_{\rm pass} > 2$}{
			STOP}
		 }
		 {Reset pass count $N_{pass}=0$     }
    }

	\caption{Adaptive CR surrogate training sampling}
\end{algorithm}

In addition, the activation functions employed in the HF surrogate in this section were modified from the previous case in which
we essentially demonstrated a proof-of-concept. Hidden layer neurons were assigned the contemporary Swish activation function~\cite{ramachandran2017searching}
\begin{align}
    \mathrm{swish}(x) = x\times \mathrm{sigmoid}(x)= \frac{x}{1 + \exp(-x)},
\end{align}
to make use of the general observations that it outperforms, or matches, the widely used ReLU activation function as a result of its improved smoothness and differentiability properties~\cite{Mercioni-swish,ramachandran2017searching}.

To map all output field quantities between [0,1], we now choose to rescale $\log_{10}(P_{\mathrm{rad}})$ between -20 and 20 using a formula similar with (\ref{eq:scale_to_1}). This then enables us
to employ an output layer of the ANN using a sigmoid activation function, noting that output variable quantities at the output layer are normalized between (0,1). This choice of activation layer was made with the intention of providing a fit-for-purpose architecture to hopefully improve training performance, but to also circumvent non-physical quantities appearing at the output layers, such as negative population densities of ions for example.
Finally, we note that the discussed changes to hyperparameters and activation functions used from the first half of this study were employed in order to enhance the fit-for-purpose of the surrogate. All subsequent ANNs were trained using the same updated hyperparameters and activation functions so that we may still focus on the differences arising due to active learning in the training routine compared to training using an arbitrarily large fixed data set.

\subsection{Experiments and discussion}
Employing the proposed modifications to the adaptive sampling routine, we present quantitative and qualitative results
of the performance of the sampling method and resulting surrogates. Here we choose to specify $Z=2$, corresponding to a deuterium and helium plasma
mixture, which is relevant to the scenario of helium dilution of a burning plasma undergoing DT fusion. All assessments in this study were conducted with Tensorflow 2.0.0 under Python 3.7.9 on a 2.9GHz Intel Core i9 CPU with 32GB of RAM.

The main change in this evolution of the adaptive sampling framework is the adjustment of the LF surrogate to now learn and predict the mapping of input fields to the MSE of the HF surrogate at the current iteration. Given the current CR problem being studied, there is no obvious \textit{a priori} choice for the value of the hyperparameter, $\lambda$, which balances exploitation and exploration of an error surface for a given input field. In our work, multiple values of $\lambda$ were trialled and tested in order to gain an empirical understanding of this hyperparameter's influence. Here, we present some results representative of the general trends observed in our study and a short discussion on this matter.

Here, we employ three values of $\lambda = 0.1, 1,$ and 10 to highlight the influence of this hyperparameter. We recall from (\ref{eq:MSE_COST}) that when $\lambda=1$ there is equal weighting between variance of the MSE (exploration) and the MSE itself (exploitation). When $\lambda > 1$ priority is placed on exploration as training data samples are collected from regions of relatively high variance in the MSE. Conversely, when $\lambda < 1$ the policy priority is on exploiting regions of high MSE itself to augment the training data set with additional samples.

Following execution of the adaptive sampling and training framework for these values of $\lambda$, given the conditions previously outlined in this section, we obtain surrogates for determining deuterium-helium plasma characteristics. The training metrics, as well as the number of required samples for reaching the training termination criteria are shown in Table~\ref{table:training}.

 \begin{table}[]
 \caption{Metrics for varying sizes of adaptively chosen training data
acquisition methods. \\}
\label{table:training}
\centering
\begin{tabular}{c|c|c|c|}
\cline{2-4}
                                                         & Training samples & \begin{tabular}[c]{@{}c@{}}Training $R^2$\end{tabular} & \begin{tabular}[c]{@{}c@{}}Training MSE\end{tabular} \\ \hline
\multicolumn{1}{|c|}{\cellcolor{red!25}Large LHS set }                      & 8000             & 0.9992                                                              & 2.6835$\times 10^{-7}$                                            \\ \hline
\multicolumn{1}{|c|}{\cellcolor{green!20}Balanced AS  }   & 655              & 0.9913                                                              & 7.0304$\times 10^{-6}$                                            \\ \hline
\multicolumn{1}{|c|}{\cellcolor{babyblueeyes}Exploitation biased AS} & 596              & 0.9962                                                              & 6.0526$\times 10^{-6}$                                            \\ \hline
\multicolumn{1}{|c|}{\cellcolor{orange!20}Exploration biased AS}  & 339              & 0.9915                                                              & 2.4972$\times 10^{-5}$                                            \\ \hline
\end{tabular}

\end{table}

As shown in Table~\ref{table:training}, using the adaptive sampling framework, with hyperparameters described previously, enables a high training validation goal of $R^2 = 0.99$ to be reached with only a few hundred samples. For a balanced policy, $\lambda=1$, 655 samples were required, compared to 596 when either exploitation or 339 when exploration were prioritized. For the raw validation metrics, the $R^2$ and MSE values for $\lambda=0.1$ are marginally the best performing. However, these metrics are not significantly better than the $R^2$ and MSE produced in the balanced $\lambda=1$ case, where both cases yield an MSE on the order of $10^{-6}$. We should note that given the capability of single precision operation of the ANN training platform, and for the context of providing fractional ion populations and a transformed RPL quantity, this error is an acceptable outcome for implementation within multiphysics plasma simulation codes. Naturally, the surrogate trained on the very large LHS dataset with 8000 samples performed very well in validation, with an order of magnitude reduction in MSE and $R^2$ of 0.999.

Following validation, given an unseen data set witheld from training and validation, we can assess the performance of the four ANN surrogates on mapping a set of input fields unseen during training. For this case, the testing $R^2$ results are shown in Table~\ref{table:testing}. For these assessments testing metrics are computed from the output fields, but in addition a $R^2$ value is computed when the output field is augmented by an additional element of the electron density, $n_e$, which can be computed in post-processing from the output field and then appended. Though the $n_e$ quantity is a derived quantity given an output field, it is a very important variable when describing a plasma, and so we include this in the assessment of our surrogates to further test their ability to reproduce physically important values.

\begin{table}[]
\caption{Testing metrics for different training data
acquisition methods.\\}
\label{table:testing}
\centering
\begin{tabular}{c|c|c|}
\cline{2-3}
                                                         & \begin{tabular}[c]{@{}c@{}}Testing\\ $R^2$\end{tabular} & \begin{tabular}[c]{@{}c@{}}Testing\\ $R^2$ w/ $n_e$\end{tabular} \\ \hline
\multicolumn{1}{|c|}{\cellcolor{red!25}Large LHS set }                      & 0.9924                                                  & 0.9931                                                           \\ \hline
\multicolumn{1}{|c|}{\cellcolor{green!20}Balanced AS  }   & 0.9923                                                  & 0.9976                                                           \\ \hline
\multicolumn{1}{|c|}{\cellcolor{babyblueeyes}Exploitation biased AS} & 0.9911                                                  & 0.9922                                                           \\ \hline
\multicolumn{1}{|c|}{\cellcolor{orange!20}Exploration biased AS}  & 0.9826                                                  & 0.9891                                                           \\ \hline
\end{tabular}

\end{table}

Here we can see that performance between the LHS trained surrogate is comparable to the performance of the adaptively trained surrogate that was exposed to over an order of magnitude less data (8000 vs.~655).  In fact, the adaptively trained case with $\lambda=1$ could be argued to slightly outperform the rest with an $R^2$ of 0.9976 compared to the surrogate trained on
the very large LHS data set. This result is encouraging, in that over an order of magnitude less sample points can be used to train an effective surrogate trained using traditional data acquisition methods.

\begin{figure}[]

		\begin{minipage}{0.5\textwidth}
		\centering
		(a)
	    \end{minipage}
	    \begin{minipage}{0.5\textwidth}
		\centering
		(b)\\
	    \end{minipage}

	    \begin{minipage}{0.5\textwidth}
		\centering
		\includegraphics[width=0.9\linewidth]{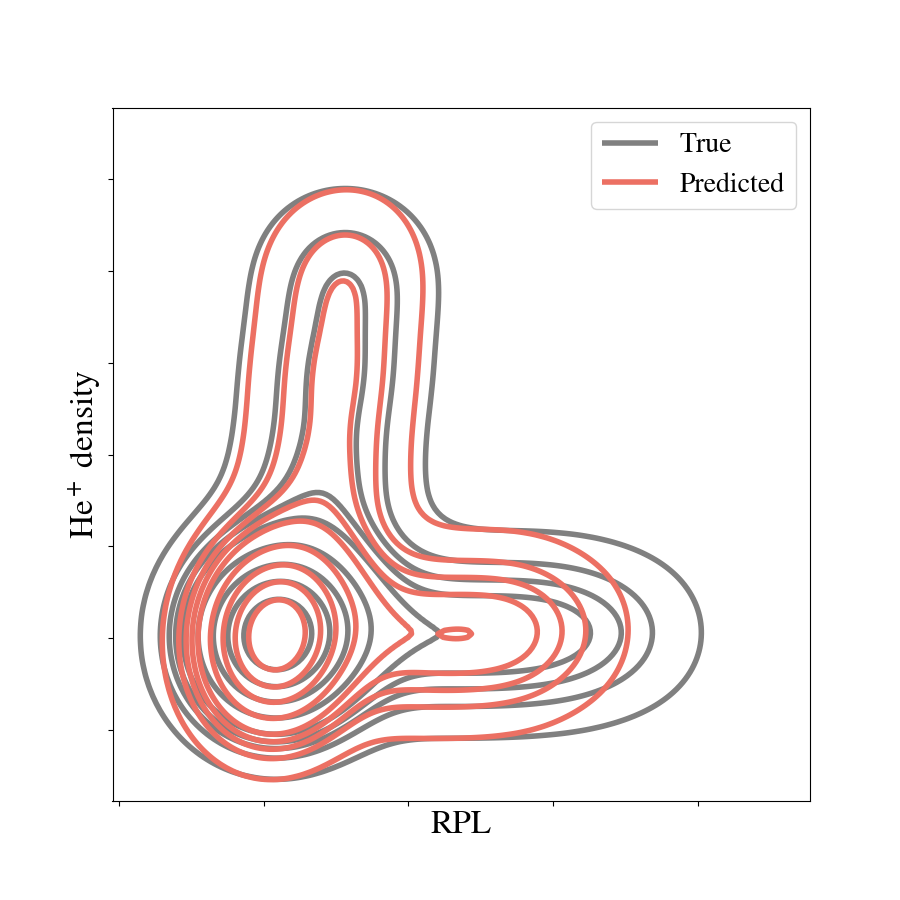}
	    \end{minipage}
	    \begin{minipage}{0.5\textwidth}
		\centering
		\includegraphics[width=0.9\linewidth]{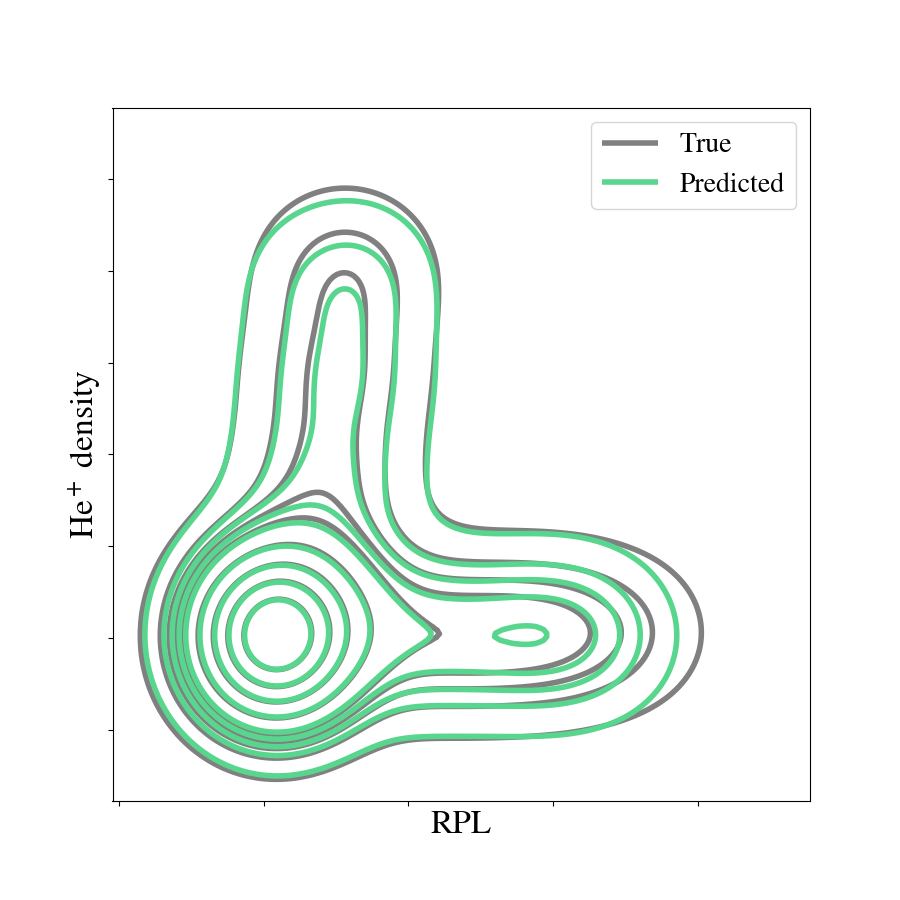}\\
	    \end{minipage}

		\begin{minipage}{0.5\textwidth}
		\centering
		(c)
	    \end{minipage}
	    \begin{minipage}{0.5\textwidth}
		\centering
		(d)\\
	    \end{minipage}

	    \begin{minipage}{0.5\textwidth}
		\centering
		\includegraphics[width=0.9\linewidth]{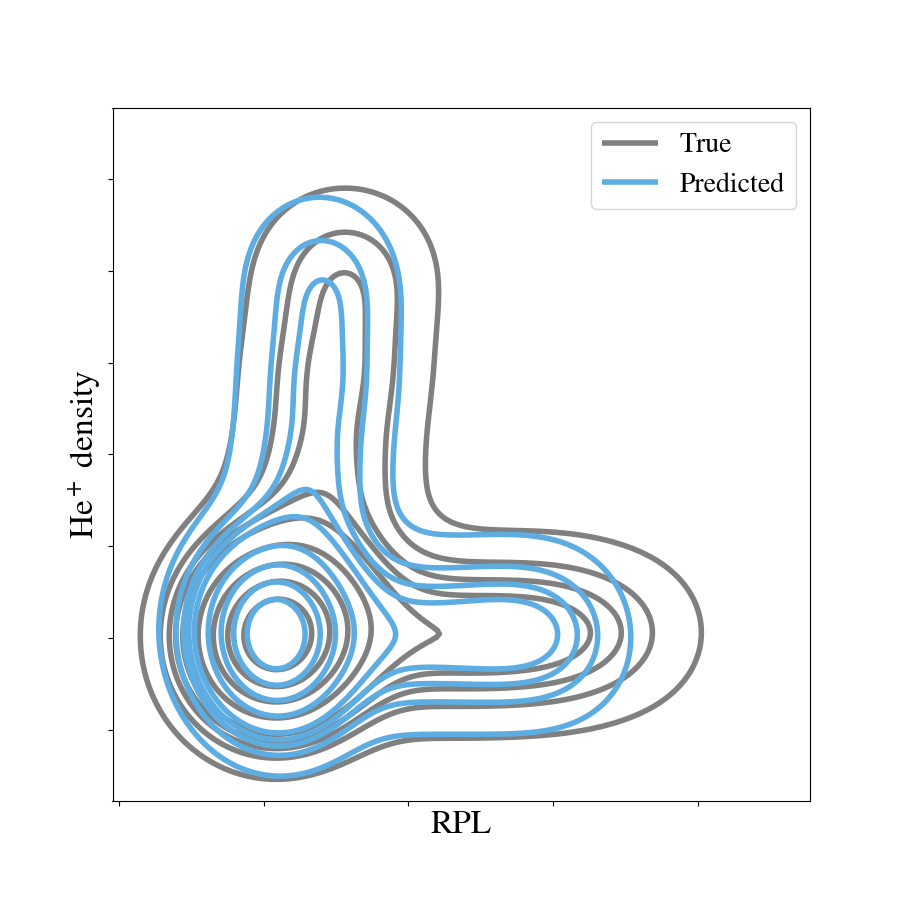}
	    \end{minipage}
	    \begin{minipage}{0.5\textwidth}
		\centering
		\includegraphics[width=0.9\linewidth]{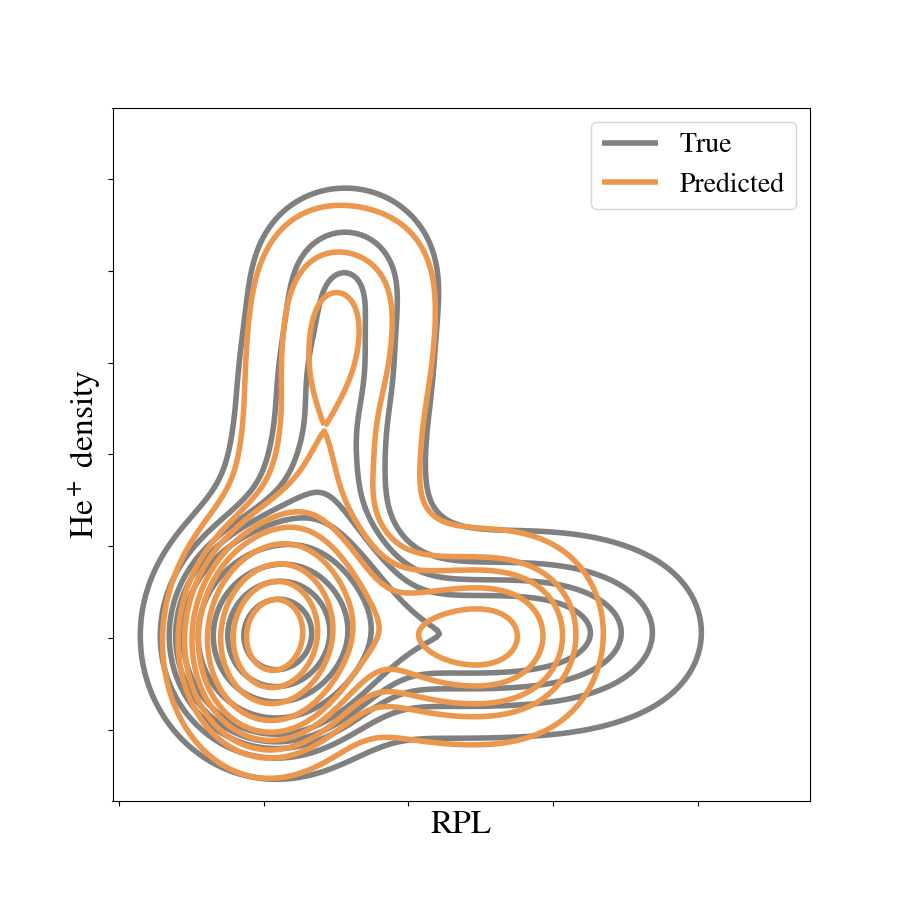}
	    \end{minipage}

	\caption{Kernel density estimate plots demonstrating the true and predicted relationship between output field quantities of total radiation $P_{\mathrm{rad}}$ and the population density of the He$^{+1}$ ion for (a) MLP trained on set of 8000 points via latin hypercube sampling, (b) MLP adaptively trained by balancing exploration with exploitation of MSE of HF surrogate (i.e. $\lambda=1$) , (c) MLP adaptively trained by weighting exploitation over exploration (i.e. $\lambda=0.1$) , and (d) MLP adaptively trained by weighting exploration over exploitation (i.e. $\lambda=10$) }
	\label{fig:PDF_e_e}
\end{figure}

While comparing $R^2$ and MSE metrics is a reasonable thing to do when evaluating surrogates, we can also visually compare the performance of potential surrogates by projecting the relationships between output field quantities onto KDE plots to provide easy graphical representation of the predicted outputs compared to the known output relationships in the testing set. These KDE plots
are shown in Figure~\ref{fig:PDF_e_e} for the  four surrogates being examined in this section.

Interestingly, comparison of subplots (a) and (b) in Figure~\ref{fig:PDF_e_e} highlights that
the MLP trained with a balance of exploitation and exploration better captures the relationship between the two output field quantities considered: He$^+$ ion density and $\log_{10}(P_{\mathrm{rad}})$ in this case. Specifically, we see that the adaptively trained surrogate is able to better predict the features in the lower half of the probability distribution than any of the other surrogates. It indicates that the predictions capable of the adaptively trained surrogate are grounded in a training set with more representative samples compared to the very large set of 8000 instances generated via LHS. This outcome is an encouraging demonstration of the intended nature of the adaptive sampling framework to identify regions of parameter space where predictions break down, and assign more sampling resources to those regions.


To provide a quantitative comparison to complement the KDE plot shown in Figure~\ref{fig:PDF_e_e}, we can compute a relative entropy measure, specifically the Kullback–Leibler divergence~\cite{shlens2014notes,cover2006elements}, between the distributions of output field quantities of a proposed surrogate and the known truth distribution. In the context of comparing an approximate distribution, $Q$, trying to emulate a true distribution, $P$, the Kullback–Leibler divergence, $D_{\mathrm{KL}}(P||Q)$, is computed as
\begin{align}\label{eq:D_KL}
    D_{\mathrm{KL}}(P||Q) = \sum_i P_i \log \frac{P_i}{Q_i},
\end{align}
and is a measure of entropy increase, or loss of information, due to the use of an approximation, $Q$, compared to the true distribution itself, $P$. When $D_{\mathrm{KL}}(P||Q)=0$ it implies that $P$ and $Q$ have identical amounts of information. Thus a smaller value of $D_{\mathrm{KL}}(P||Q)$ indicates a better surrogate performance. For the RPL and He$^+$ ion density variables visualized in Figure~\ref{fig:PDF_e_e}
we show $D_{\mathrm{KL}}$ values in Table~\ref{table:DKL}.
The table shows that the improved visual similarity observed in the KDE plots when adaptively sampling with a balanced policy in exploitation and exploration is indeed reflected in the relative entropy measures between the candidate surrogates.

\begin{table}[]
\caption{Kullback-Leibler divergence measures for radiation loss and He$^+$ ion density variables visualized in Figure~\ref{fig:PDF_e_e}.\\ }
\label{table:DKL}
\centering
\begin{tabular}{|c|c|c|}
\hline
\textbf{Sampling model} & \textbf{$D_{\mathrm{KL}}$ for RPL} & \textbf{$D_{\mathrm{KL}}$ for $n_e$} \\ \hline
\cellcolor{red!25}Large LHS set           & $5.13 \times 10^{-3}$              & $5.28 \times 10^{-4}$                \\ \hline
\cellcolor{green!20}Balanced AS             & $3.03 \times 10^{-3}$              & $2.81 \times 10^{-4}$                \\ \hline
\cellcolor{babyblueeyes}Exploitation biased AS  & $5.81 \times 10^{-3}$              & $1.66 \times 10^{-3}$                \\ \hline
\cellcolor{orange!20}Exploration biased AS   & $1.19 \times 10^{-2}$              & $1.40 \times 10^{-1}$                \\ \hline
\end{tabular}
\end{table}

\begin{figure}[]

		\begin{minipage}{0.45\textwidth}
		\centering
		(a)
	    \end{minipage}
	    \begin{minipage}{0.45\textwidth}
		\centering
		(b)
	    \end{minipage}
	    \begin{minipage}{0.45\textwidth}
		\centering
		\includegraphics[width=0.9\linewidth]{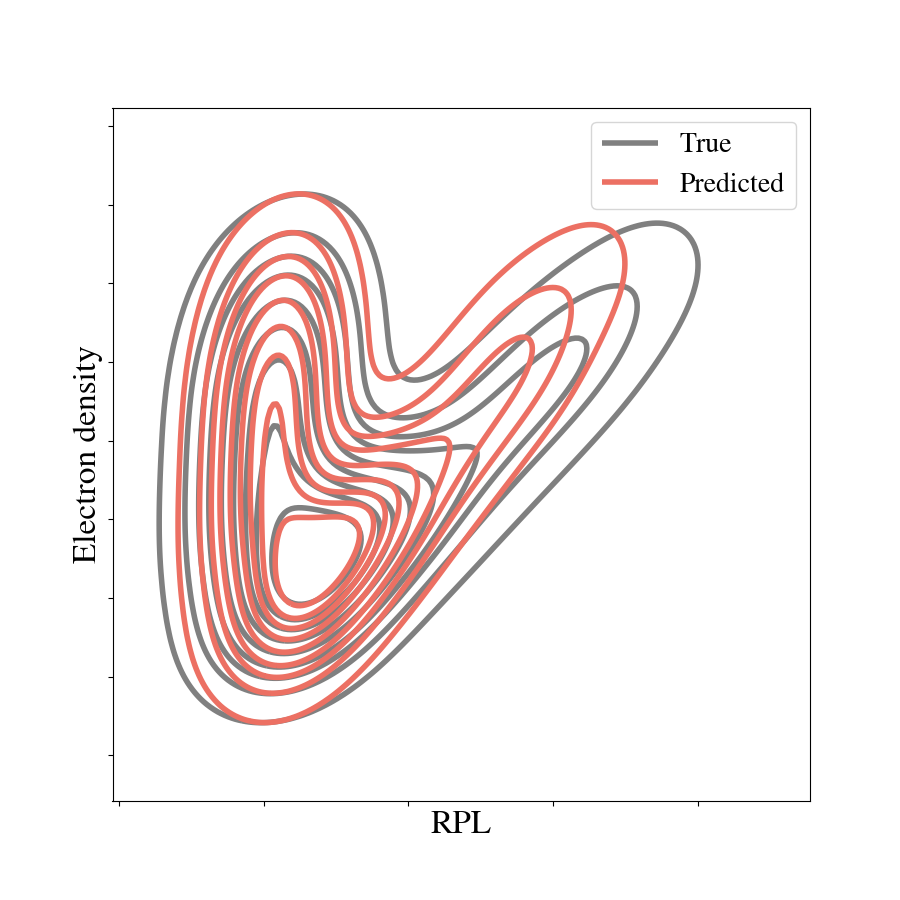}
	    \end{minipage}
	    \begin{minipage}{0.45\textwidth}
		\centering
		\includegraphics[width=0.9\linewidth]{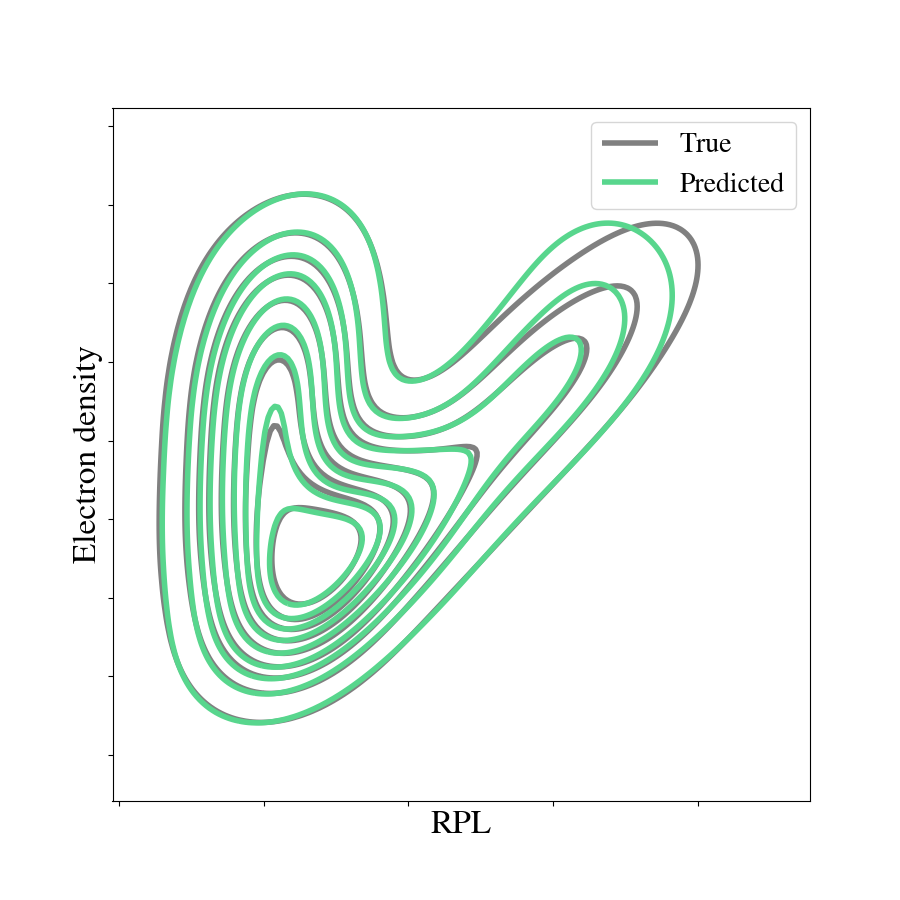}
	    \end{minipage}

		\begin{minipage}{0.45\textwidth}
		\centering
		(c)
	    \end{minipage}
	    \begin{minipage}{0.45\textwidth}
		\centering
		(d)
	    \end{minipage}
	    \begin{minipage}{0.45\textwidth}
		\centering
		\includegraphics[width=0.9\linewidth]{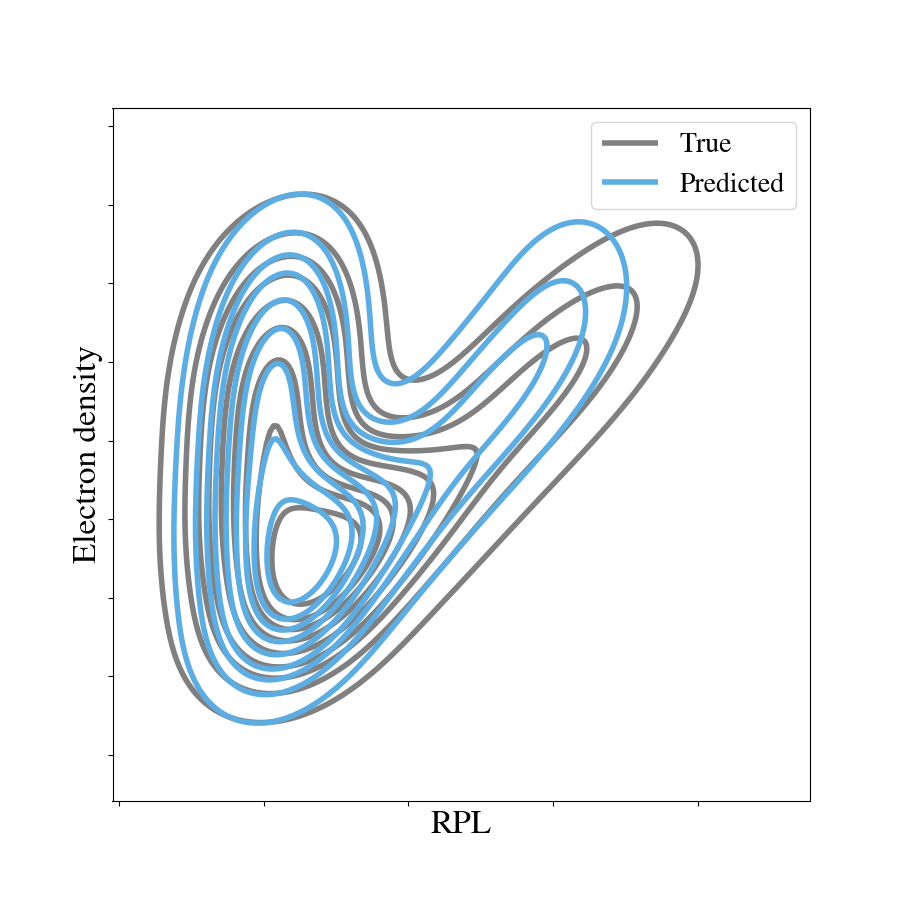}
	    \end{minipage}
	    \begin{minipage}{0.45\textwidth}
		\centering
		\includegraphics[width=0.9\linewidth]{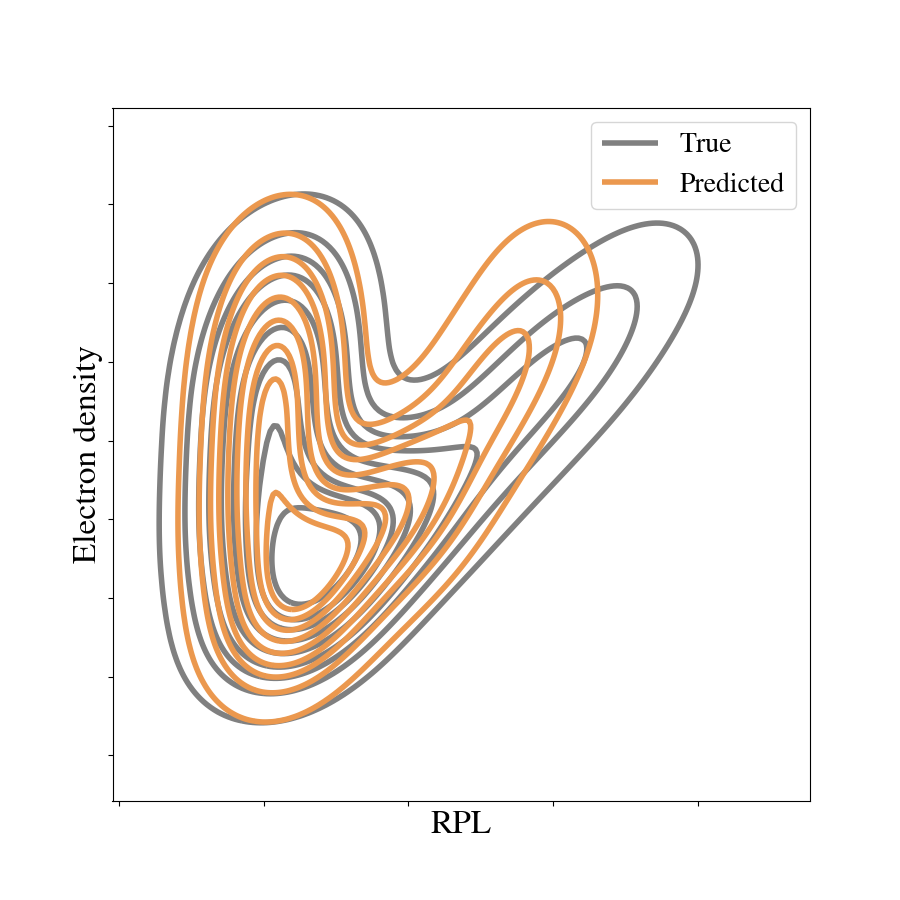}
	    \end{minipage}
	\caption{Kernel density estimate plots demonstrating the true and predicted relationship between output field quantities of total radiation $P_{\mathrm{rad}}$ and the density of electrons for (a) MLP trained on set of 8000 points via latin hypercube sampling, (b) MLP adaptively trained by balancing exploration with exploitation of MSE of HF surrogate (i.e. $\lambda=1$) , (c) MLP adaptively trained by weighting exploitation over exploration (i.e. $\lambda=0.1$) , and (d) MLP adaptively trained by weighting exploration over exploitation (i.e. $\lambda=10$) }
	\label{fig:PDF_ne}
\end{figure}

\begin{figure}[]

        \begin{minipage}{0.23\textwidth}
		\centering
		(a)
	    \end{minipage}
	    \begin{minipage}{0.23\textwidth}
		\centering
		(b)
	    \end{minipage}
	    \begin{minipage}{0.23\textwidth}
		\centering
		(c)
	    \end{minipage}
	    \begin{minipage}{0.23\textwidth}
		\centering
		(d)
	    \end{minipage}

	    \begin{minipage}{0.23\textwidth}
		\centering
		\includegraphics[width=0.96\linewidth]{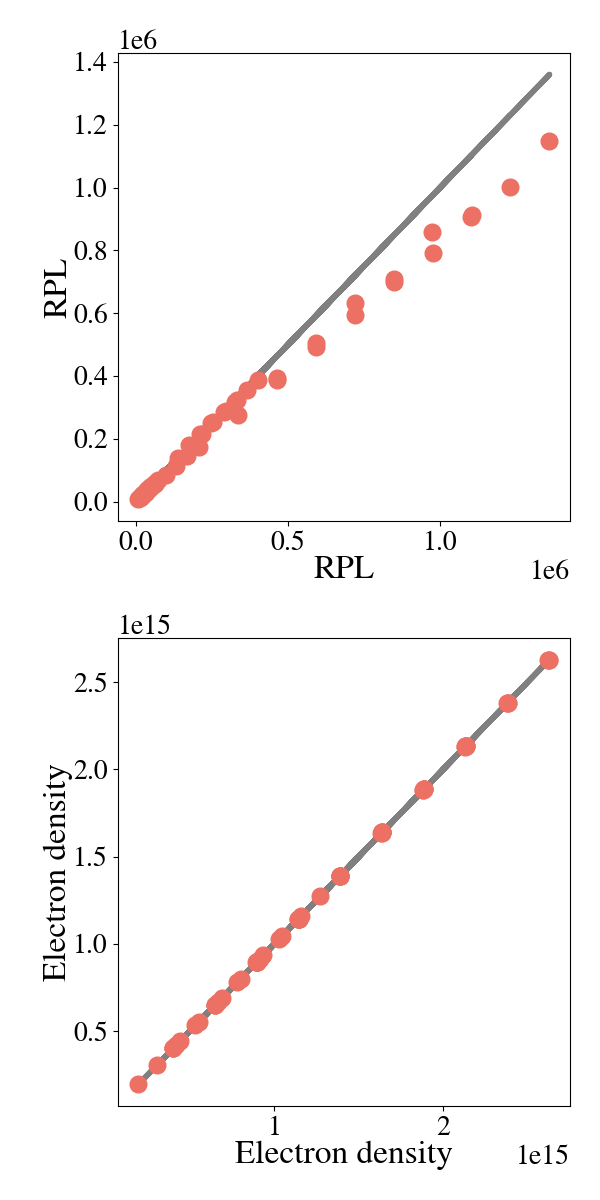}
	    \end{minipage}
	    \begin{minipage}{0.23\textwidth}
		\centering
		\includegraphics[width=0.96\linewidth]{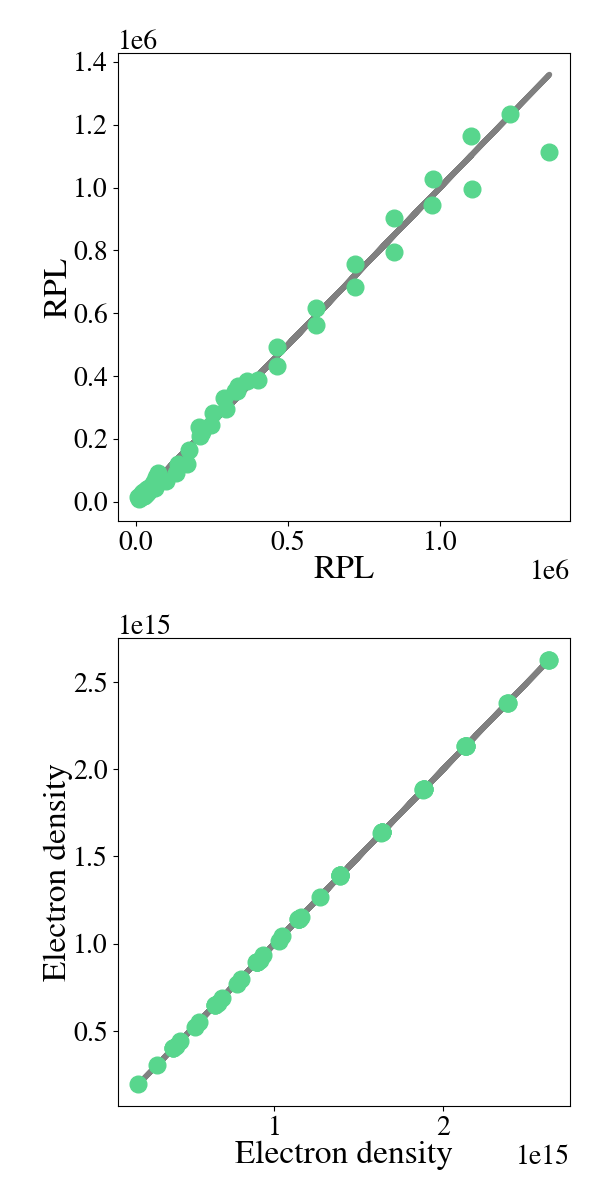}
	    \end{minipage}
	    \begin{minipage}{0.23\textwidth}
		\centering
		\includegraphics[width=0.96\linewidth]{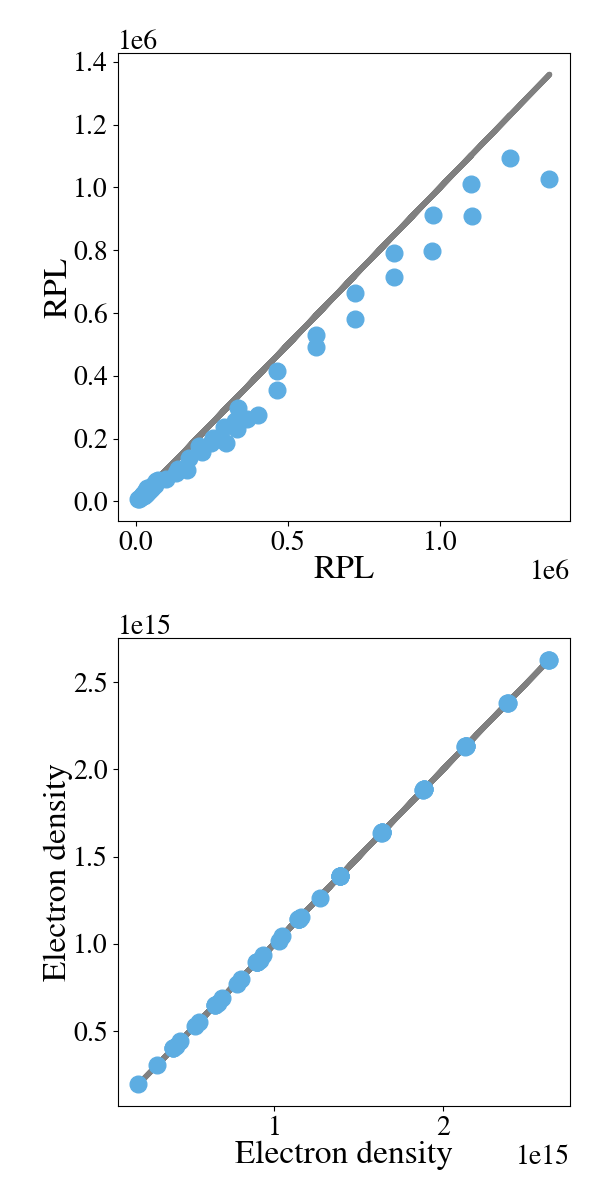}
	    \end{minipage}
	    \begin{minipage}{0.23\textwidth}
		\centering
		\includegraphics[width=0.96\linewidth]{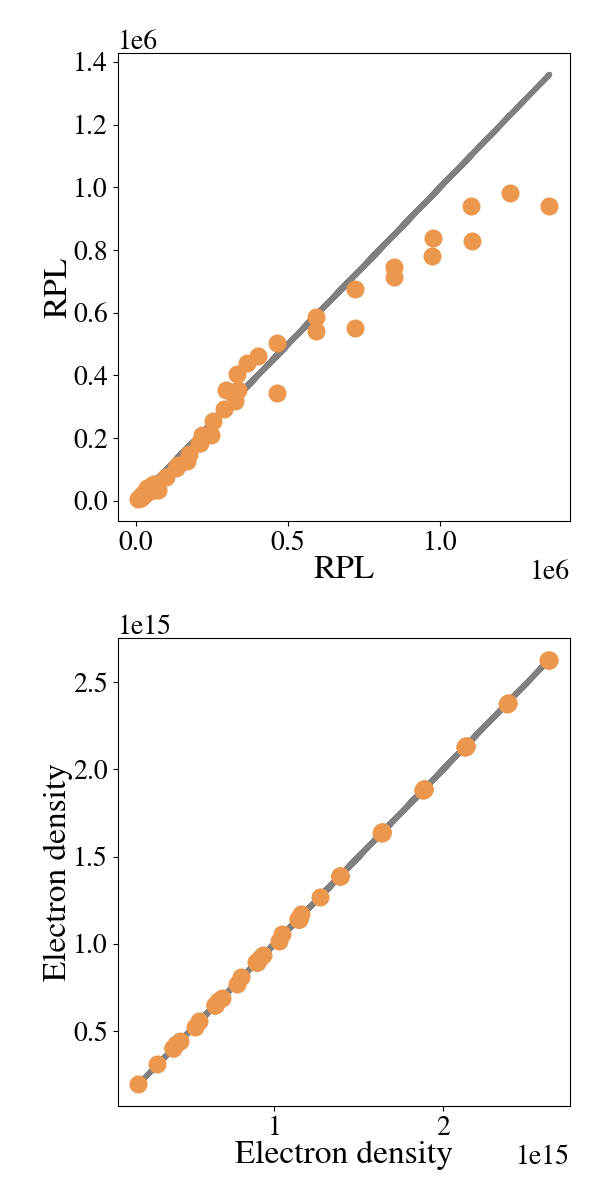}
	    \end{minipage}

	\caption{Scatter plots demonstrating the true and predicted relationship between output field quantities of (top) total radiation, $P_{\mathrm{rad}}$, and (bottom) electron density, $n_e$, for (a) MLP trained on set of 8000 points via latin hypercube sampling, (b) MLP adaptively trained by balancing exploration with exploitation of MSE of HF surrogate (i.e. $\lambda=1$) , (c) MLP adaptively trained by weighting exploitation over exploration (i.e. $\lambda=0.1$) , and (d) MLP adaptively trained by weighting exploration over exploitation (i.e. $\lambda=10$) }
	\label{fig:scatter_ne}
\end{figure}

As reinforcement of the observations made from the previous KDE visualization, we present a second KDE figure showing the relationship between electron density and RPL, which are arguably the two most physically meaningful quantities in plasma characterization, in Figure~\ref{fig:PDF_ne}.
A common visualization to assess ANN performance is a scatter graph of predicted output field values against known true values. For the important RPL and electron density variables used in Figure~\ref{fig:PDF_ne}, we present their scatter plots in Figure~\ref{fig:scatter_ne}. Figure~\ref{fig:scatter_ne} shows that all surrogates appear to do a reasonable job at predicting the electron density for a set of unseen input fields. However, as suggested in the KDE plots, the predictive capability for RPL is less than ideal for most surrogates, except the adaptively trained surrogate using $\lambda=1$.

It is not surprising that RPL is a more challenging quantity to reproduce than electron density, or ion charge state populations, given the well known physics of the coronal equilibrium limit~\cite{bauche2015atomic,ralchenko2016modern}. In the coronal limit plasma densities are assumed to be relatively low, and ion populations are assumed to be in the ground state. Ion population balance is determined between ionization and recombination between ground states of ion charge states, and so very simple models can replace full CR models to determine charge state populations. Where this physical limit breaks down is in the determination of excited states and thus subsequent RPL quantities, in intermediate regimes where arbitrary populations of ion excited states can occur. This CR regime is where our current tokamak plasmas appear to sit, though in some scenarios accurate ion charge state populations can be determined from coronal-like models if the conditions are appropriate~\cite{whyte_disruption_2003}. Ultimately, from these physics considerations, the total RPL quantity is very sensitive to small changes in excited state populations, and thus makes for a more challenging quantity to learn accurate mappings for when compared to ion charge state populations.

Despite the reduced size of the training data set, this surrogate is able to predict the RPL much better than the alternatives. Again, this result underscores the virtue of the adaptive sampling framework to not only produce a sufficiently accurate surrogate in a shorter time, with fewer training samples required, but it also ensures a physically meaningful set of data is gathered. By exposing the MLP training infrastructure to a much more representative set of samples we can then take greater confidence that a surrogate from this method can be used to bypass an original forward pass CR model coupled within a multiphysics plasma simulation. Consequently, with confidence in the fidelity of a CR surrogate, that can be demonstrably produced cheaply, there are now greater opportunities for greatly improved coupling between atomic and plasma physics models at next to no computational expense thanks to the nature of rapidly executable ANN surrogates.

\begin{figure}[]
	\centering
	\includegraphics[width=0.6\linewidth]{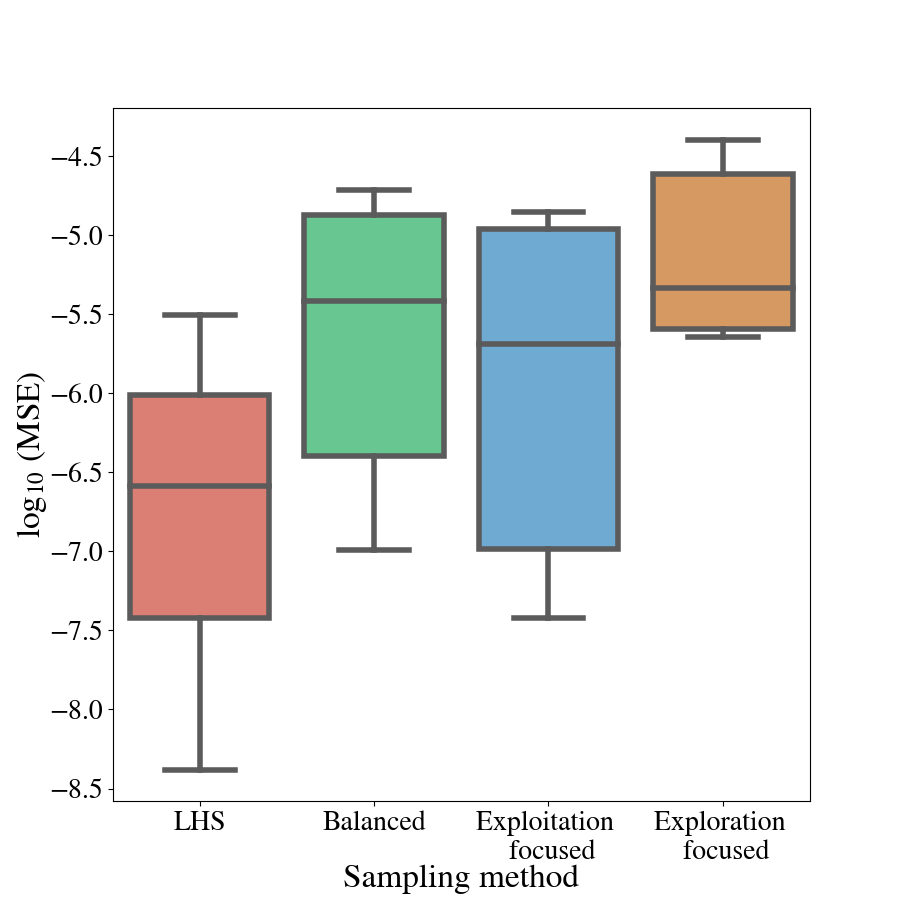}
	\caption{Box plot of testing mean squared errors over output layer
	fields (a) MLP trained on set of 8000 points via latin hypercube sampling, (b) MLP adaptively trained by balancing exploration with exploitation of MSE of HF surrogate (i.e. $\lambda=1$) , (c) MLP adaptively trained by weighting exploitation over exploration (i.e. $\lambda=0.1$) , and (d) MLP adaptively trained by weighting exploration over exploitation (i.e. $\lambda=10$) }
	\label{fig:mse_box}
\end{figure}

Finally, we now comment on a measurement in which the large LHS data set remains superior: the MSE, as shown in Table~\ref{table:training}, where a $\mathcal{O}(10^{-7})$ training MSE was reported compared to the $\mathcal{O}(10^{-6})$ training MSE for the cheaper balanced adaptively sampled surrogate. To briefly explore this quantity, we present box plots of the testing MSEs obtained over output layer fields in Figure~\ref{fig:mse_box}. Here, we can see that broadly the ANN surrogate trained on the LHS set is able to maintain a window of approximately an order of magnitude around a MSE  $~\mathcal{O}(10^{-6.5})$. The balanced adaptively sampled ANN surrogate is able to maintain a similar bandwidth of MSE but is pinned at a higher $~\mathcal{O}(10^{-5.5})$ MSE.

We note that while not the best overall candidate in our study, the exploitation-biased adaptively-sampled ANN surrogate was able to yield a lower MSE floor than the balanced sampling case.  The latter was able to provide results which were more aligned with the true behavior when comparing KL-divergence metrics. This outcome is a direct result of the objective function seeking to minimize the surrogate MSE in training. Another demonstration of the impact of our objective function policy can be seen in the MSE of the exploration-biased surrogate. Here, the bandwidth of MSE error is by far the smallest due to the penalty added to minimize variance of the MSE, but not the MSE itself. As a result, the MSE floor and ceiling are higher for this model despite having a narrow spread. Demonstration of these impacts on MSE due to objective function policy is an encouraging result, and will allow further tailoring and improvements on surrogate training depending on potential MSE requirements of a surrogate's application. Furthermore, we may also use a combination of metrics to determine the sampling strategy for new data sets.


\section{Conclusion and Future Work}\label{sec:summary}
In this study we have outlined the motivation for employing ANN-based surrogates of CR models in coupled multiphysics plasma codes, so that vital quantities can be provided to the physics simulation in a computationally efficient manner, avoiding a bottleneck due to the costly CR models. We have demonstrated the capability of an adaptive sampling framework to autonomously generate
training data in statistically meaningful areas of the problem parameter space, which in turn allows: (i) faster acquisition of data samples for training/validating the surrogate, and (ii) some assurance that the data assembled for training/validation is physically meaningful and representative of the problem parameter space. Further, we have shown that applications of this framework towards creation of an ANN-based surrogate for a CR model, that would allow rapid evaluations of ionization balance and radiation properties within a fusion plasma doped with an impurity species, outperform training
an ANN-based surrogate on more than 10 times the amount of data generated by the  often used LHS technique. Future research will explore transfer learning of the ANN (rather than retraining), and parallelized sampling, for example employing DeepHyper \cite{maulik2020recurrent}, as well as implementation of the surrogates presented in this study into plasma simulation research codes. Another direction will be to apply the adaptive sampling framework outlined in this study to more physics rich, but expensive, forward CR models.

\section*{Acknowledgements}
This work was supported by the U.S. Department of Energy (DOE), Office of Science, Office of Advanced Scientific Computing Research, under Contract No. DE-AC02–06CH11357, at Argonne National Laboratory, and by the Office of Fusion Energy Sciences and Office of Advanced Scientific Computing Research under the Scientific Discovery through Advanced Computing (SciDAC) project of Tokamak Disruption Simulation at Los Alamos National Laboratory (Contract No. 89233218CNA000001). Partial support was also provided by the Office of Fusion Energy Sciences under the DeepFusion pilot project in scientific machine learning and artificial intelligence for fusion energy sciences. This research was funded in part by the US DOE Laboratory Directed Research and Development (LDRD) program, under grant number 20200356ER. This research was funded in part and used resources of the Argonne Leadership Computing Facility, which is a DOE Office of Science User Facility supported under Contract No. DE-AC02–06CH11357. We acknowledge funding support from ASCR for DOE-FOA-2493 “Data-intensive scientific machine learning”.



\newcommand{\newblock}{}
\newpage
\bibliographystyle{unsrt}
\bibliography{Atomic,Plasma,ML}

\end{document}